\documentclass[preprint,12pt]{elsarticle}

\usepackage{amssymb}
\usepackage{comment}
\usepackage{relsize}
\usepackage{libertinust1math}

\usepackage{svg}
\usepackage{amsmath}
\usepackage{dsfont}
\usepackage{color,soul}

\journal{Physica A: Statistical Mechanics and its Applications}

\begin{document}

\begin{frontmatter}



\title{Towards effective information content assessment: analytical derivation of information loss in the reconstruction of random fields with model uncertainty}


\author[inst1]{Aleksei Cherkasov}

\affiliation[inst1]{organization={Moscow Institute of Physics and Technology},
            addressline={Institutskiy Pereulok 9}, 
            city={Dolgoprudny},
            postcode={141700}, 
            country={Russia}}

\author[inst2]{Kirill M. Gerke\corref{maincorr}}
\cortext[maincorr]{Corresponding author}
\ead{kg@ifz.ru}
\author[inst1,inst2]{Aleksey Khlyupin}

\affiliation[inst2]{organization={Schmidt Institute of Physics of the Earth of Russian Academy of Sciences},
            addressline={Bolshaya Gruzinskaya 10}, 
            city={Moscow},
            postcode={123242}, 
            country={Russia}}
\begin{abstract}

Structures are abundant in both natural and human-made environments and usually studied in the form of images or scattering patterns. To characterize structures 
a huge variety of descriptors is available spanning from porosity to radial and correlation functions. In addition to morphological structural analysis, 
such descriptors are necessary for stochastic reconstructions, stationarity and representativity analysis. The most important characteristic of any such 
descriptor is its information content - or its ability to describe the structure at hand. For example, from crystallography it is well known that experimentally 
measurable $S_2$ correlation function lacks necessary information content to describe majority of structures. The information content of this function can be assessed
using Monte-Carlo methods only for very small 2D images due to computational expenses. Some indirect quantitative approaches for this and other correlation function were also proposed. Yet, to date no methodology to obtain information content for arbitrary 2/3D image is available. In this work, we make a step toward developing a general framework
to perform such computations analytically. We show, that one can assess the entropy of a perturbed random field and that stochastic perturbation of fields’ correlation function decreases its information content. In addition to analytical expression, we demonstrate that different regions of correlation function are in different extent informative and sensitive for perturbation. Proposed model bridges the gap between descriptor-based heterogeneous media reconstruction and information theory and opens way for computationally
effective way to compute information content of any descriptor as applied to arbitrary structure.

\end{abstract}

\begin{graphicalabstract}
\includegraphics[width = \textwidth]{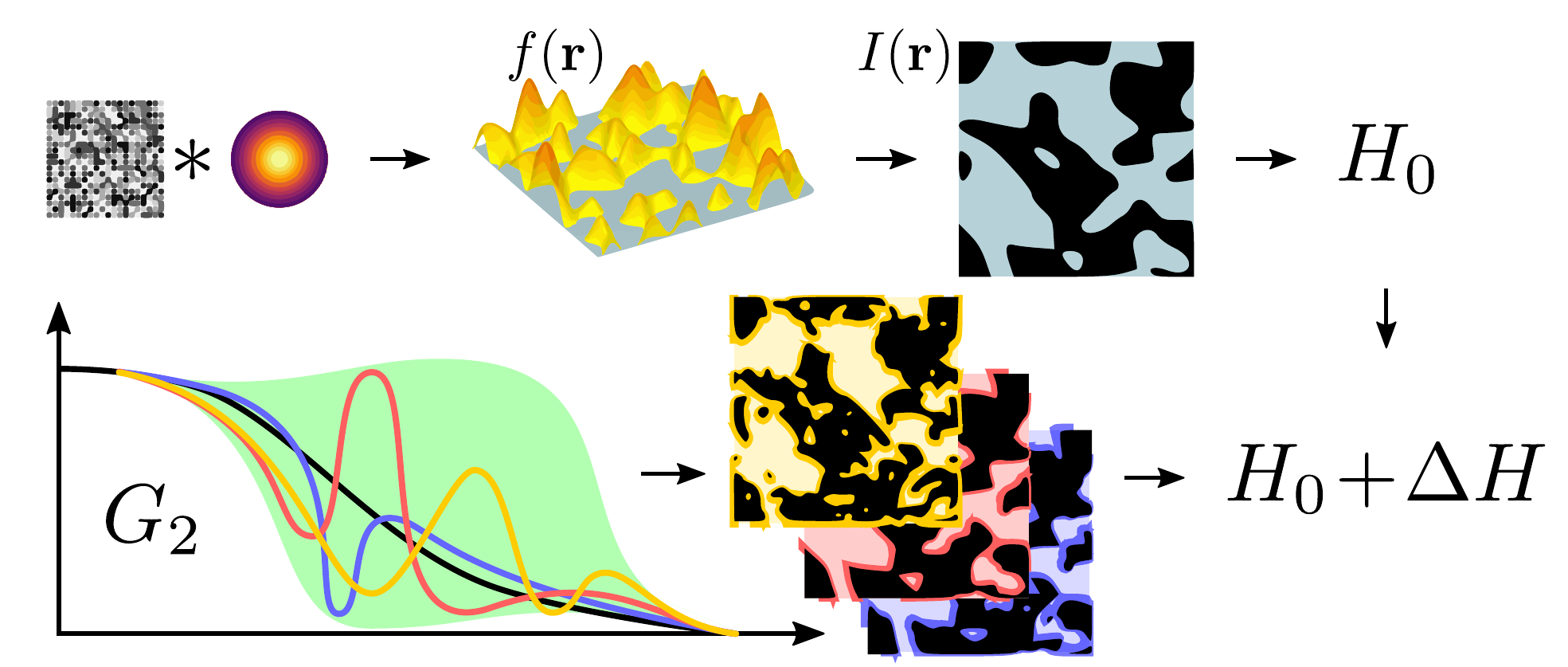}
\end{graphicalabstract}

\begin{highlights}
\item Shannon entropy for random fields with model uncertainty can be explicitly expressed by developed theoretical model 
\item Stochastic perturbation of fields' correlation function decreases its information content
\item Different regions of correlation function are in different extent informative and sensitive for perturbation
\item Proposed model bridges the gap between descriptor-based heterogeneous media reconstruction and information theory
\end{highlights}

\begin{keyword}
correlation functions \sep structure characterization \sep structural descriptors \sep image analysis \sep information content
\end{keyword}

\end{frontmatter}


\section{Introduction}
\label{sec:intro}

Structures, or spatial arrangements of some matter or phases, are ubiquitous in nature, and, thus, a target of studies in numerous research areas. Examples may include galaxy formations \cite{springel2006}, immiscible multi-phase fluid flow patterns \cite{hopkins2015new,balashov2021}, rock and soil samples \cite{rozenbaum2014,karsanina2015,ledesma2018,chen2020super,prokhorov2021digital}, food specimens \cite{derossi2019,nagdalian2021} or biological tissue \cite{park2020}. The scale of interest may span orders of magnitude from as small as couple of nm for nanoporous materials \cite{garum2020,gerke2021} to millions of light-years for star clusters \cite{hopkins2013stars}. Studied structures are not necessarily static and may exhibit temporal variation in their arrangement \cite{jiao2013,fomin2023soil} which can be a result of interacting with surrounding material. Closed or open, one needs to characterize such systems of structures in order to assess their physical properties \cite{gerke2019tensor,rozanski2023} - an ultimate goal needed to describe complex behavior of studied systems with numerical models. A wide varieties of descriptors were developed in order to quantify spatial structure, for example: phase ratio or porosity, surface area, radial distribution functions \cite{zimm1948scattering,becker2010radial}, Minkowski functionals \cite{vogel2010} and tensors \cite{schroder2011minkowski}, correlation functions \cite{Torquato_book} among numerous others]. Holy wars on the question which metric is superior still abound e.g.,\cite{vogel2022holistic,yudina2023dual}. In reality, this issue boils down to a single characteristic.

The universal characteristic of any descriptor for structures is information content - the measure of the extend the given metric describes given structure. The challenge that explains the above-mentioned diversity of structural metrics is, however, the complexity in its evaluation. But let us start from the simplest example of porosity for a binary structure containing only pores and solid material. The information content in this case can be computed as:
\begin{gather}
I_{porosity}= N-log_{2}(\substack{N\\N_{pores}}).
\label{eq:kirill1}
\end{gather}
This represents a very simple idea that if we have a binary image and know its size then $I=-log_{2}(p)=N$ (any realization of the structure is evenly distributed with the probability of $p=1/N^2$), where $N$ is the total number of pixels (i.e., the size of the image) and logarithm insures that information content is expressed in bits (i.e., one bit per pixel). Now if in addition to that we know the porosity (phase ratio of pore to solid pixels on the image), we simply obtain Eq.\ref{eq:kirill1} by substituting the amount of information contained in specifying the value of $N_{pores}$ \cite{gommes2012pre}, as opposed to knowing only the system size. Here we immediately come to a number of important observations: 1) the information content of porosity alone is rather small (which explains the doomed to fail quest to predict physical properties such as permeability from porosity values \cite{nur1998critical,chapuis2003use} and very poor performance of Kozeny-Carman like relationships \cite{kozeny1927,carman1937}) for anything different from sphere packings in general, and 2) it varies with the value of porosity with the maximum at 0.5 (like porosity any metric may change its content with small alterations in the structure itself). But before we move on to some more involved examples, the notion of stochastic reconstructions is in order.

The classical example of information content in action is the so-called crystallography problem. The standard correlation function $S_2$ can be measured from small angle scattering experiments \cite{debye1957scattering} and it can be used to stochastically reconstruct the studied structure. A variety of techniques can be applied to recover the structure based on its autocorrelation \cite{gommes2018stochastic} among which truncated random Gaussian fields was one of the pioneering approaches \cite{adler1990flow}. It was immediately recognized that $S_2$ alone is not enough to recover the measured structure. Gaussian random field, for example, can produce numerous realizations of the same auto-correlation function that are all valid solutions. Such solutions are called degenerate states. As later became clear with the progress in stochastic reconstruction methods, the problem is even deeper- not only we are not aware which realization is the target one, but stochastic reconstructions preferentially converge to a limited amount of solutions \cite{EPL2,vcapek2018importance}. Interestingly, the latter is also true to reconstruction techniques not based on correlation functions which include multiple-point statistics \cite{tahmasebiPRL,feng2018accelerating,gravey2020quicksampling} and deep learning methods \cite{zhang2022improved,volkhonskiy2022generative}. To summarize, autocorrelation, while much more informative than porosity, is not enough to describe any but the simplest of structures \cite{gommes2012pre}; one needs to increase the information content of metrics used for stochastic reconstructions in order to improve their accuracy.

Increasing the information content of structural metrics is convoluted, as it involves a trade-off between the amount of information and complexity of computation combined with the size of the metric. It is well recognized that $n$-point correlation function completely describes any structure. However, it has basically the size of structural image itself. For example, full correlation map of 2-point statistics is known to be enough to perform the stochastic reconstruction exactly \cite{chubb2000every,fullwood2008}, but its size is four-times the original for 2D image. As adepts of correlation functions, we shall use them in this work; moreover, we argue that compared to other metrics they possess a number of important qualities: 1) in addition to sampling from images some correlation functions can be measured from experiments \cite{debye1957scattering,dietrich1995scattering,li2018direct}; 2) a set of classical correlation functions \cite{Torquato_book} contains all conventional metrics \cite{karsanina2021compressing}, 3) unlike mainly scalar metrics, correlation functions can be computed in directions \cite{jiao2014modeling,gerke2014improving} and, thus, account for structural anisotropy, 4) the order of correlation functions is useful measure of their information potential with full correlation map serving as 100\% reference. This does not mean that information content can not be established for other metrics - it does and Eq.\ref{eq:kirill1} provides an example. Multiple-point statistics produces results resembling the original structure better if the size of the training window is enlarged (i.e., $n$-pointness is increased). It was also demonstrated that increasing the order of functions from 2-point to $n$-point have diminishing returns with each step \cite{yao1993high,gommes2012pre}. The trick is to utilize light-weight metrics that describe different aspects of structure. In practice this means using different correlation functions with low order, for example, $S_2$, $L_2$ and two-point surface functions \cite{vcapek2011transport,karsanina2018enhancing,adam2022efficient}. It was shown that the usage of cluster $C_2$ functions adds a lot of information, but its usage in reconstructions \cite{jiao2009superior} is cumbersome due to computationally ineffective optimization in its recalculations for annealing updates. While considerable speed gains were achieved with hierarchical annealing \cite{campaigne2012frozen,karsanina2018hierarchical}, what is the best order and number of correlation functions that should be used to reconstruct the structure at hand is still an open question. All this highlights the importance of solving the information content problem.

The straightforward way to assess information content is enumerate all possible structures that all satisfy the same correlation function or a set of correlation functions. This can be achieved by exploring the energy landscape of the stochastic reconstruction problem \cite{wang2001efficient} and was applied to $S_2$ correlation function \cite{gommes2012prl} in a similar manner we analyzed all possible variations for a given porosity value and image size in Eq.\ref{eq:kirill1}. As we shall analyze later on, this is a very computationally expensive procedure that prohibits its usage for arbitrary structure and set of functions. Other indirect approaches allow only qualitative estimates based on reconstructions \cite{chen2020probing} or analysis of specific Debye-type of structures \cite{skolnick2021understanding}. In other words, we are still missing a universal and computationally effective approach to quantify information content. Information content evaluation is needed not only for stochastic reconstructions. Physical properties of any material depend on its structure, the boundary conditions \cite{gerke2019tensor,thovert2020influence,scandelli2022computation,chen2022impacts} and representativeness. The latter quality is tricky, but in many cases crucial to evaluate \cite{zhang2000pore,gerke2021pore,ghanbarian2022estimating}. Strictly speaking, only statistically homogeneous structures can be representative. Choosing descriptors with sufficient information content for (in)homogeneity characterization is also important for stochastic reconstructions \cite{tahmasebi2015,gommes2009,karsanina2023stochastic}, structure compression \cite{SciRep,Havelka} and extracting features for machine learning \cite{kamrava2020linking,roding2020predicting,cheng2022data}. In order to establish spatial stationarity one needs high information content \cite{EfimEPL}, as otherwise the degenerate states may render homogeneity to be established incorrectly (false positive cases of stationarity \cite{gerke2021pore}).

In light of the overall introductory discussion it seems that different structural descriptors have appeared in wide variety of research disciplines due to historical and ease of measuring reasons. To tame this motley of concepts and visions requires solving the problem of efficient information content evaluation for arbitrary structures. In this light, the aim of this paper is to develop an analytical approach to compute entropy for digital image after some perturbation to its descriptors is introduced. We believe this achievement to be a crucial step towards computing information content for any arbitrary structural descriptor at hand.


\section{Theoretical Model}
\label{sec:entropy}
We consider random fields $\textbf{f}$ that can be sampled knowing the underlying probability model $P\left(\textbf{f}\right)$. Since the possible value of the field at each space point runs through a series of continuous values, then, strictly speaking, the number of different realizations of random fields $\textbf{f}$ is infinite. Therefore, the Shannon entropy \cite{shannon1948mathematical} of such random fields is of interest as a measure of the information contained in their probabilistic model $P\left(\textbf{f}\right)$. 

More specifically, our goal is the investigation of the uncertainty introduced into the underlying probabilistic model. If we add a random perturbation $U$ to any parameter $\theta$ that determine the probability $P\left(\textbf{f}; \theta + U\right)$, then the entropy growth of original fields $\textbf{f}$ is expected. This is pretty clear, since we have lost some information while adding such uncertainty which obeys its own probability law $\rho(U)$. Thus, in this paper, we are interested in a quantitative estimate of such an additional Shannon entropy introduced into the system due to uncertainty.

In our model Gaussian fields $\textbf{f}$ are considered with a given arbitrary correlation matrix $\hat{G_2}$. A random symmetric matrix $\hat{U}$ added to the original  matrix $\hat{G_2}$ acts as a small perturbation in the model. We show that one of the simple ways to investigate the additional entropy is to apply an orthogonal decomposition of random fields known as the Karhunen--Loève transform \cite{hristopulos2020random}. By combining this approach with matrix perturbation theory the desired expression for the entropy can be obtained, which is demonstrated below during step by step calculations. Our theoretical evaluation consists of three main steps. Firstly,  we evaluate joint probability in the case of small perturbation of certain correlation matrix in the basis of correlation matrix eigenvectors. After that we provide the perturbation approach for Shannon information entropy using previous evaluations. Finally, desired expression for additional entropy is obtained making use of some averaging over random matrix probability model.

\subsection{The Karhunen--Loève Transform for Random Fields}
\label{subsec:th}
Arbitrary random field can be decomposed into a series in terms of an orthogonal functional basis using  the Karhunen--Loève transform. For any finite number of terms this decomposition is the best in the sense of Euclidean norm of difference between the field and its approximation \cite{hristopulos2020random}. KLT and its discrete version (also commonly known as principal component analysis - PCA) have a wide range of applications such as bioinformatics \cite{ma2011principal}, \cite{stacklies2007pcamethods}, data analysis \cite{mishra2017multivariate}, dimensionality reduction methods for direct \cite{brunton2016discovering}, \cite{carlberg2018conservative} and inverse problems \cite{sarma2006efficient}, \cite{elizarev2021objective} of high computational complexity, applications in quantum physics and quantum information theory \cite{vladimirov2019quantum}.

It will be convenient to provide our further perturbation analysis based on the KLT of random fields. So here we briefly introduce the KLT in the case of the non-perturbed correlation function $G_2$ which characterises a certain ensemble of Gaussian random fields $\mathbf{f}(\mathbf{r})$. The correlation function $G_2$ given by
\begin{gather}
G_2\left(\textbf{x}\right)=G_2\left(|\textbf{x}|\right)=\left\langle f(\textbf{r}) f(\textbf{r}+\textbf{x}) \right\rangle
\end{gather}
corresponds to the unperturbed correlation matrix 
\begin{gather}
(\hat{G_2})_{ij}=G_2(\textbf{r}_\textbf{i}- \textbf{r}_\textbf{j})=G_2(|\textbf{r}_\textbf{i}- \textbf{r}_\textbf{j}|)
\end{gather}

The set of correlated variables $\textbf{f} = [f(\textbf{r}_1), \dots, f(\textbf{r}_n)]^T$ has the joint probability density 
\begin{gather}
\rho\left(\textbf{f}\right) = \frac{|\hat{G_2}|^{-\frac{1}{2}}}{(2\pi)^{\frac{n}{2}}}\exp\left(-\frac{1}{2}\textbf{f}^T\hat{G_2}^{-1}\textbf{f}\right)
\end{gather}

Due to the Karhunen--Loève expansion a certain realization of such a field can be presented in the eigenfunction $\psi_k$ basis as
\begin{gather}
f(\textbf{r}) = \sum_{k = 1}^n\psi_k(\textbf{r})\chi_k
\end{gather}
with independent variables
$\chi_k$ normally distributed with probability density
\begin{gather}
\rho_k= \frac{1}{\sqrt{2\pi\lambda_k}}e^{-\frac{\chi_k^2}{2\lambda_k}} \quad \forall k \in 1,...,n
\end{gather}

As stated earlier, $\psi_k$ are the eigenfunctions of the correlation matrix $\hat{G_2}$ and $\lambda_k$ are their eigenvalues.
\begin{gather}
\hat{G_2} \psi_k = \lambda_k \psi_k \quad \forall k \in 1,...,n
\end{gather}

The joint probability density of the whole realization $\boldsymbol{\chi} = (\chi_1, \dots \chi_n)$ in this ortogonal space of eigenfunctions $\psi_k$ is now the product of independent probability densities for  $\chi_k$
\begin{gather}
\label{eq:joprob}
\rho\left(\boldsymbol{\chi}\right) =\prod_{k=1}^n\rho_k\left(\chi_k, \lambda_k\right)
\end{gather}

\subsection{A Random Field with an Uncertainity of the Correlation Function}
\label{subsec:pert}
Since the joint probability density (\ref{eq:joprob}) depends on the eigenvalues $\lambda_k$ it changes when the correlation matrix is perturbed. In this section, we use perturbation theory to modify the expression (\ref{eq:joprob}) regardless of the stochastic nature of the perturbation $\hat{U}$ of the correlation matrix \cite{hirschfelder1964recent,landau2013quantum} Further, we will take into account the explicit form of this perturbation in section \ref{sec:excess}. 

The perturbed correlation function
$G_2(x) + U(x)$ corresponds to the correlation matrix $\hat{G_2} + \hat{U}$ with eigenvectors $\phi_k$ and eigenvalues $\beta_k$
\begin{gather}
\left[\hat{G_2} + \hat{U}\right]\phi_k = \beta_k\phi_k \quad \forall k \in 1,...,n
\end{gather}

The modified eigenvalues $\beta_k = \lambda_k + \lambda_{k,1} + \lambda_{k,2}$ now include first- and second-order corrections \cite{landau2013quantum}:
\begin{gather}
\label{eq:lam2}
\lambda_{k,1} =P_{kk} \\
\lambda_{k,2} = \sum_{m \neq k}\frac{P_{km} P_{mk}}{\lambda_k - \lambda_m}\nonumber
\end{gather}
where the matrix element $P_{km}$ can be considered as a representation of $\hat{U}$ in the basis set $\psi_k$
\begin{gather}
\label{eq:pkm}
P_{km} = (\psi^k, \hat{U} \psi^m )=\sum_{i,j}\psi^k_i\hat{U}_{ij} \psi^m_j
\end{gather}

The probability density perturbation 
can be estimated using the second-order Taylor expansion w.r.t perturbation $\hat{U}$
\begin{gather}
\label{eq:dens_full}
\rho_k(\chi_k| \hat{G} + \hat{U})
= \frac{1}{\sqrt{2\pi(\lambda_{k} + \lambda_{k,1} + \lambda_{k,2})} }\exp\left[-\frac{\chi_k^2}{2(\lambda_{k} + \lambda_{k,1} + \lambda_{k,2})}\right]= \\
=\rho_k(\chi_k, \lambda_k)\left(1 + {f}_k(\chi_k, \hat{U})\right) \nonumber
\end{gather}

The correction ${f_k} = A(\hat{U}) + B(\hat{U})\chi_k^2 + C(\hat{U})\chi_k^4$ is a quadratic polynomial of $\chi_k^2$. The coefficients $A$, $B$ and $C$ contain terms of the  first and second order on perturbation $\hat{U}$ (since eigenvalue corrections functionally depend on $\hat{U}$)
\begin{gather}
A = \frac{1}{\lambda_k}\left( - \frac{1}{2}\lambda_{k,1}- \frac{1}{2}\lambda_{k,2}+  \frac{3}{8}\lambda_{k,1}^2 \right)\\ \nonumber
B = \frac{1}{2\lambda_{k}^2}\left({\lambda_{k,1}} + {\lambda_{k,2}} -  \frac{3}{2}{\lambda^2_{k,1}}\right)\\
C = \frac{1}{8\lambda_{k}^3}{\lambda^2_{k,1}} \nonumber
\end{gather}

 We give a detailed evaluation of the probability density and even expressions for coefficients $A$, $B$ and $C$ in \ref{ap:1}.
From this point on, we denote $\rho_k(\chi_k| \hat{G_2} + \hat{U})$ as $\rho_k(\chi_k| \hat{U})$ for simplicity and bear in mind that this corresponds to a certain $\hat{G_2}$.
Thus conditional joint probability density for the whole set of $\chi_i$ now contains the corrections $f_k$
\begin{gather}
\label{eq:fuldens}
\rho(\boldsymbol{\chi}|\hat{U}) = \prod_{k=1}^n\rho_k(\chi_k, \lambda_k)\left(1 + {f}_k(\chi_k, \hat{U})\right)
\end{gather}

\subsection{Excess Shannon Entropy}
\label{sec:excess}
In the section \ref{subsec:pert} we derived the joint probability density independently of the random nature of the perturbation $\hat{U}$. In this section, we impose a number of conditions on the perturbation and evaluate an expression for Shannon entropy of the field with a perturbed correlation matrix for the certain probability model. Starting from the  unperturbed Shannon entropy \cite{shannon1948mathematical}
 \begin{gather}
 \label{eqn:entr0}
H_0 = -\int d \boldsymbol{\chi}  \sum_i \ln(\rho_{i})\prod_k \rho_{k}
\end{gather} 
one may obtain the expected known result for Gaussian multivariate entropy:
\begin{gather}
\label{eqn:entr_unp}
H_0 = -\sum_i \int d\chi_i \rho_{i} \ln(\rho_{i})
=\frac{1}{2}\ln\left[(2\pi e)^n\det(\hat{G_2})\right]
\end{gather}

\begin{figure*}
  \centering
  \includegraphics[width = 0.6\textwidth]{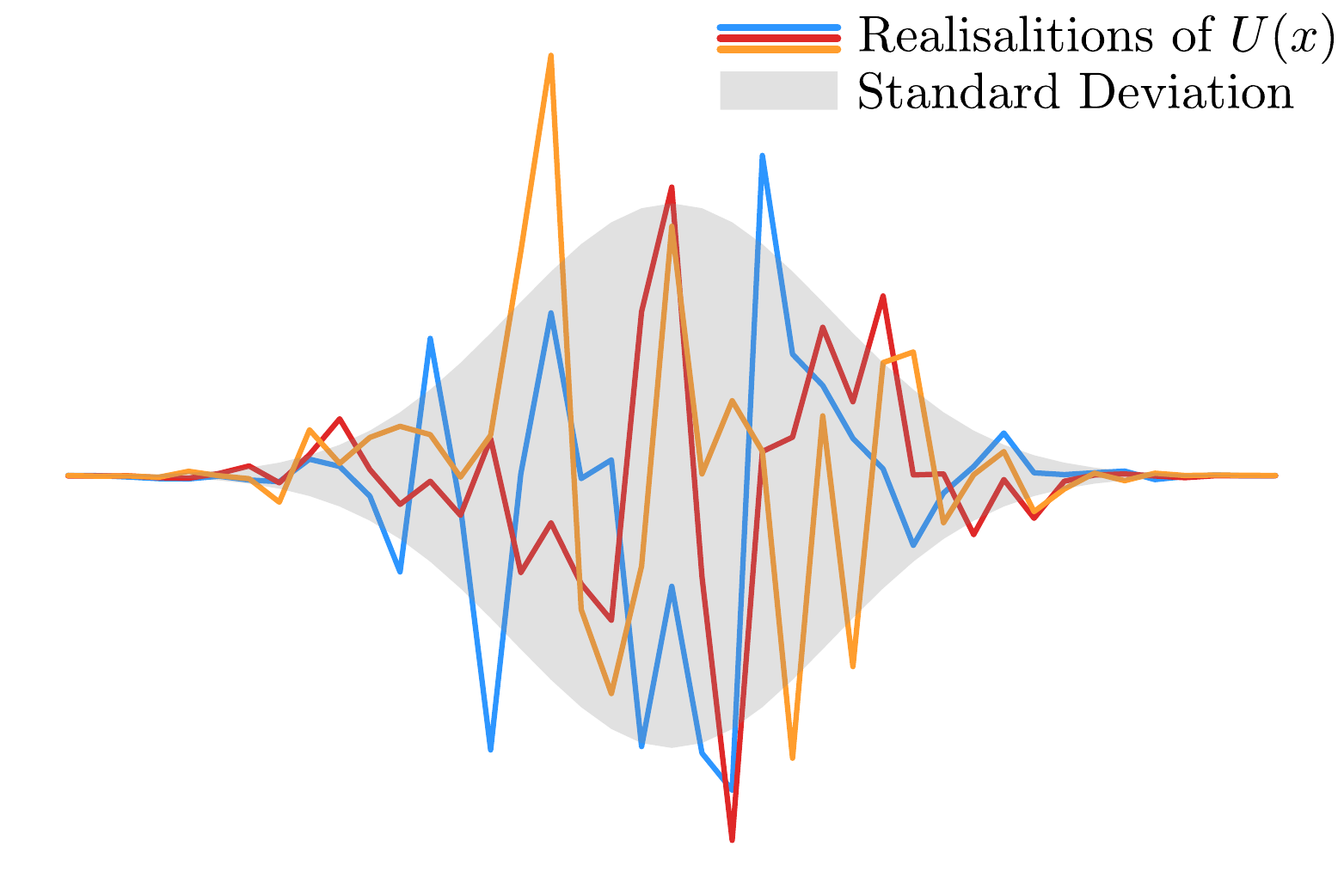}
  \caption{Three different realizations of the perturbation $U(x)$ from the distribution with zero mean and smooth variance.}
    \label{fig:pert_scheme}
\end{figure*}

 Now we move to perturbed conditional distribution and make several assumptions that allow us to obtain an analytical result for the entropy of such a stochastic perturbation $\hat{U}$.
First, we assume that the perturbation has a zero mean, and its variance is a smooth function $u(x)$. Different realisations of perturbation are completely uncorrelated (see figure \ref{fig:pert_scheme})
\begin{gather}
\label{eq:evcov}
\mathds{E}[U(x)] = 0\\
    \langle U(x) U(x+\tau)\rangle=u(x)\delta(\tau)\nonumber
\end{gather}

Slight perturbations of correlation function leads to slight perturbation of resulting field which results in loss of information content. Now we express the change of Shannon entropy due to perturbation of correlation function.

The expression for the perturbed Shannon entropy is defined as
\begin{gather}
\label{eqn:ent}
H = -\mathds{E}_{\boldsymbol{\chi}}\ln\left\langle\rho(\boldsymbol{\chi}|\hat{G_2} + \hat{U})\right\rangle_{\hat{U}} = \\\nonumber
=-\int d\boldsymbol{\chi}\int \rho(\chi|\hat{G_2} + \hat{U})\rho(\hat{U})d\hat{U}
 \ln\left(\int \rho(\chi|\hat{G_2} + \hat{U})\rho(\hat{U})d\hat{U}\right)
 \end{gather}

We integrate the conditional probability density over various realizations $U(x)$ of perturbation leading to probability density no longer depends on a certain realization of perturbation. We take only terms up to second order of perturbations:
\begin{gather}
\label{eqn:rho}
\int\rho(\chi|\hat{U})\rho(\hat{U})d\hat{U}=\langle\prod_k\rho_k\left(1 +  f_k \right)\rangle_{\hat{U}}
    =\left\langle1 + \sum_i{f_i}+\frac{1}{2}\sum_{i\neq j}{f_i}{f_j}\right\rangle_{\hat{U}}\prod_k\rho_k =\\ 
    =\left(1 + \sum_i\tilde{f_i}+\frac{1}{2}\sum_{i\neq j}\langle{f_i}{f_j}\rangle_{\hat{U}}\right)\prod_k\rho_k\nonumber
\end{gather}

Let us denote 
\begin{gather}
    \tilde{f_i} = \langle f_i \rangle_{\hat{U}} = \left\langle A + B\chi_i^2 + C\chi_i^4 \right\rangle_{\hat{U}}=\langle A \rangle_{\hat{U}} + \langle B \rangle_{\hat{U}} \chi_i^2 + \langle C \rangle_{\hat{U}} \chi_i^4 
\end{gather}

 The logarithm of the probability density from (\ref{eqn:ent}) averaged over $\hat{U}$  also contains the above-mentioned corrections $\tilde{f_i}$ and $\sum_{i\neq j}\langle{f_i}{f_j}\rangle_{\hat{U}}$
\begin{gather}
\label{eqn:lnrho}
\ln\left\langle\rho(\chi,\hat{U})\right\rangle_{\hat{U}} = \sum_k \ln \rho_k + \sum_i \tilde{f_i}  +\sum_{i\neq j}\langle{f_i}{f_j}\rangle_{\hat{U}}
\end{gather}

Now we can obtain the expression for the perturbed entropy (\ref{eqn:ent}) using (\ref{eqn:rho}, \ref{eqn:lnrho})
\begin{gather}
\label{eqn:ent_raw}
H =-\int d\chi \prod_k\rho_k\left(1 + \sum_i \tilde{f_i}+\sum_{i\neq j}\langle{f_i}{f_j}\rangle_{\hat{U}}\right) \Bigg(\sum_j\ln(\rho_j) + \sum_l \tilde{f_j}\\ \nonumber+\sum_{i\neq j}\langle{f_i}{f_j}\rangle_{\hat{U}}\Bigg) =\\
= H_0 + H_1 + H_2 \nonumber
\end{gather}

We group the terms containing $\tilde{f_i}$ and $\langle{f_i}{f_j}\rangle_{\hat{U}}$ into the summands $H_1$ and $H_2$, respectively
\begin{gather}
    H_1 = - \int d\boldsymbol{\chi}\sum_i \tilde{f}_{i}\prod_k \rho_{k} - 
\int d \boldsymbol{\chi} \sum_i\ln(\rho_{i})\sum_j \tilde{f_j}\prod_k \rho_{k}\\
H_2 = - \int d\boldsymbol{\chi} \sum_{i\neq j}\langle{f_i}{f_j}\rangle_{\hat{U}}\prod_k \rho_{k} - 
\int d \boldsymbol{\chi} \sum_i\ln(\rho_{i})\sum_{i\neq j}\langle{f_i}{f_j}\rangle_{\hat{U}}\prod_k \rho_{k}
\end{gather}

\begin{figure*}[htbp]
  \centering
  \includegraphics[width = 0.9\textwidth]{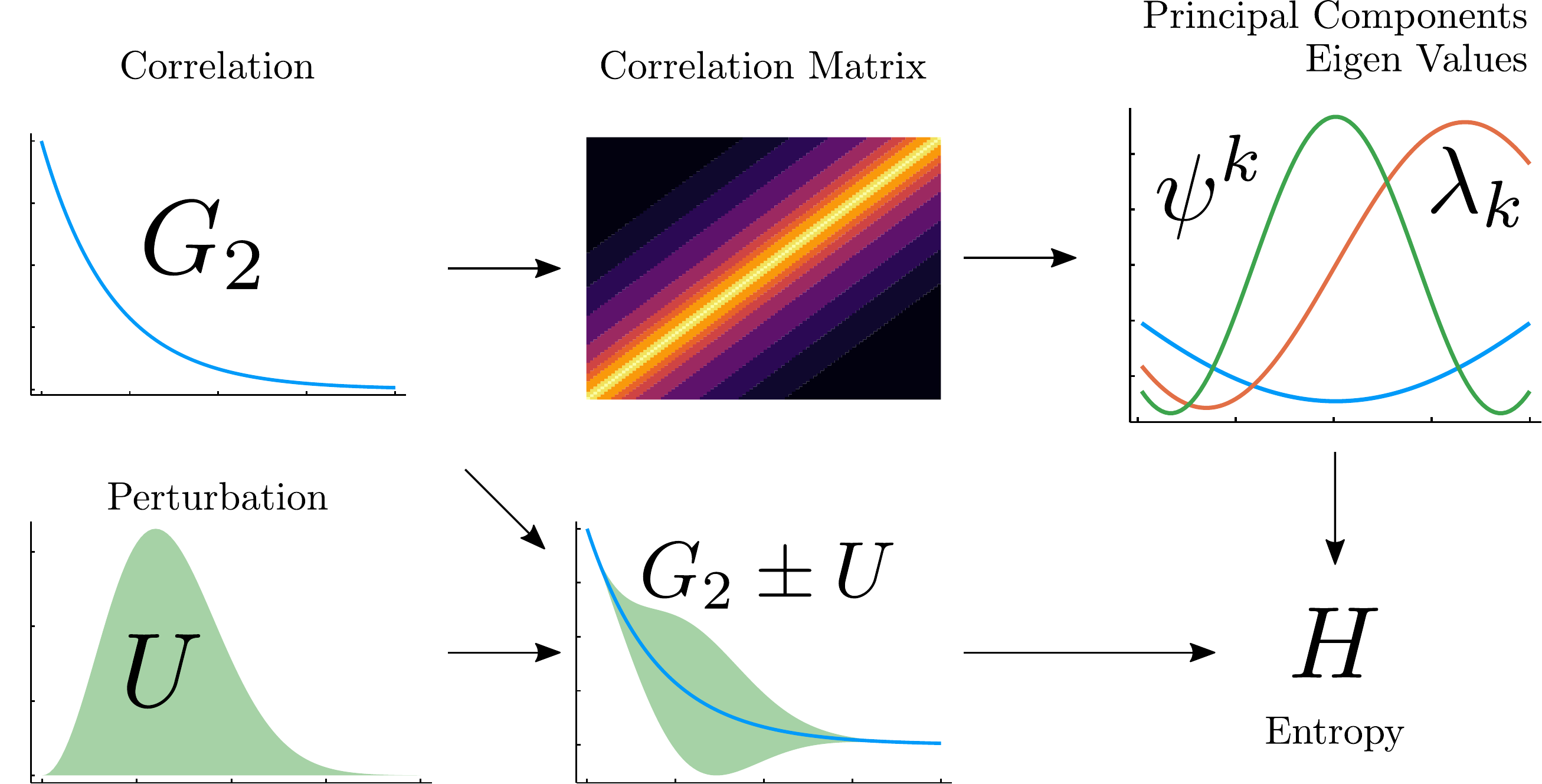}
  \caption{Stages of entropy calculation: calculate eigenvectors and principal components of correlation matrix for certain autocorrelation function, obtain entropy value for certain perturbation}
    \label{fig:scheme}
\end{figure*}

The terms containing $\tilde{f_i}\tilde{f_j}$, $\tilde{f_k}\langle{f_i}{f_j}\rangle_{\hat{U}}$, $\langle{f_i}{f_j}\rangle_{\hat{U}}\langle{f_k}{f_l}\rangle_{\hat{U}}$ are fourth-order infinitesimals, so we neglect them. The final expression for entropy is as follows
\begin{gather}
\label{eqn:ent_sfin}
    H=H_0 +\frac{1}{2}\sum_k \left(\frac{\langle\lambda_{k,1}\rangle_{\hat{U}}}{\lambda_{k}} + \frac{\langle\lambda_{k,2}\rangle_{\hat{U}}}{\lambda_{k}}\right)
\end{gather}

Thus, the change in entropy with a small perturbation of the correlation function is (see Figure \ref{fig:scheme} for schematic illustration):
\begin{gather}
\label{eq:delta_h}
    \Delta H = H - H_0= \sum_{k,m} \frac{1}{2\lambda_{k}(\lambda_k - \lambda_m)}{\sum_{l,n}(\psi^k_l)^2u_{ln}(\psi^m_n)^2}
\end{gather}

Evaluation of the previous expression is technically complex so we give detailed calculations in \ref{ap:2}.  There we also show that $H_2$ does not contribute to entropy.
\subsection{Computational Workflow and Model Application}
\begin{figure}
  \centering
  \includegraphics[width = 0.9\linewidth]{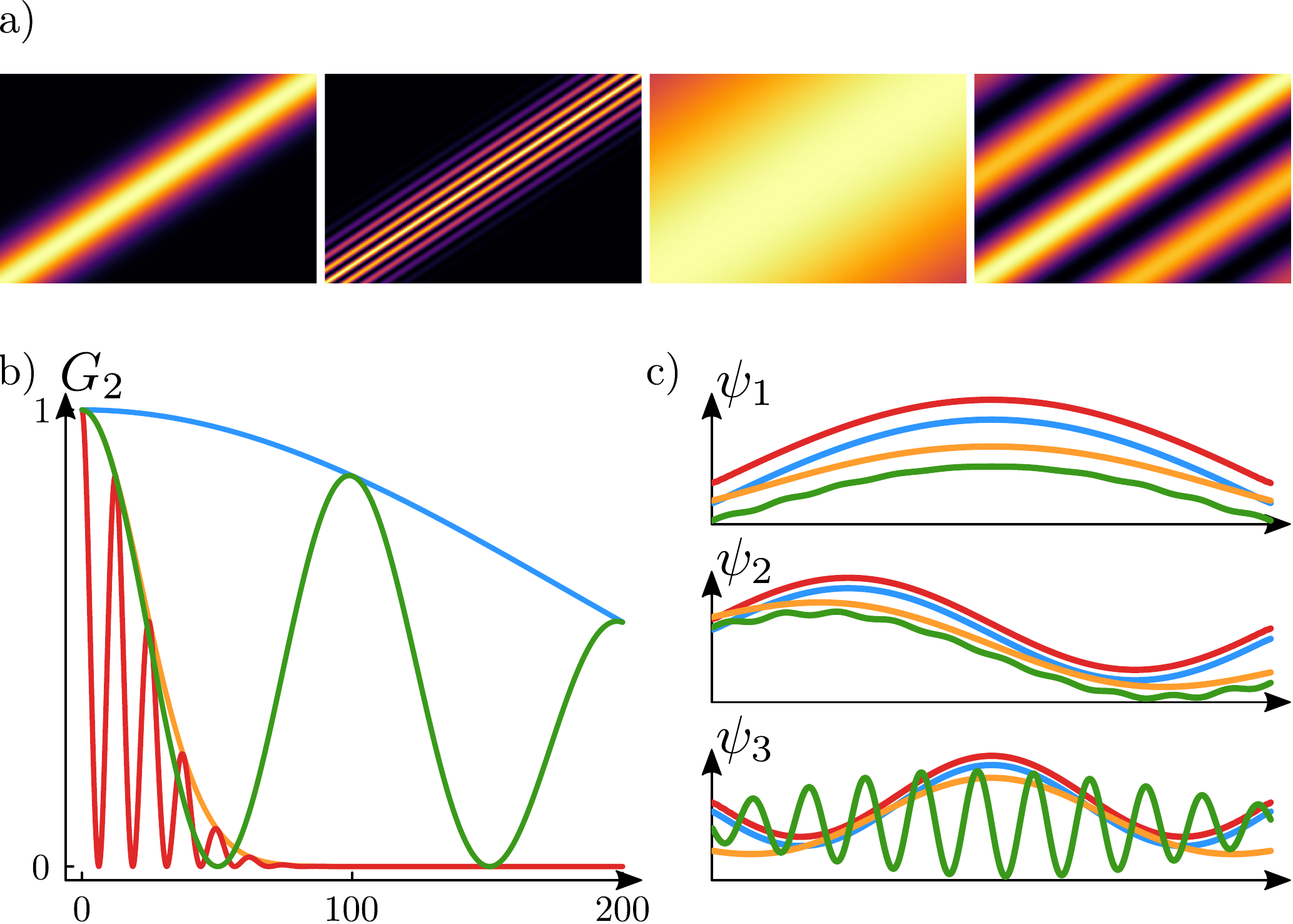}  
  \caption{a) correlation matrices; b) Monotonous and oscillating correlation functions; c) Principal components of correlation matrices. Color characterizes one from four correlation functions}
  \label{fig:corfs}
\end{figure}

We utilize the following workflow (Figure \ref{fig:scheme}) to illustrate our analytical expression for entropy. First, we compute the correlation matrix based on the given correlation function of the Gaussian field. Then, using singular value decomposition \cite{golub2013matrix} we obtain a set of eigenvalues and eigenvectors for the correlation matrix $G_2$. Finally, we apply the expression (\ref{eq:delta_h}) to calculate the desired additional entropy. We apply this procedure to Gaussian correlation functions of two different widths and to cosine modulated Gaussian functions (Figure \ref{fig:corfs}b). The corresponding correlation matrices are shown in the Figure \ref{fig:corfs}a and Figure \ref{fig:corfs}c shows first three principal components $\psi^k$ of these matrices. They are in a way similar to harmonic functions and follow a certain trend - more "narrow" $G_2$ is, the $\psi^k$ oscillates more frequently.

\subsection{Entropy Analysis with Several Uncertainty Models}
To analyse the proposed model we consider four different parameterized perturbations and study the entropy behaviour depending on their parameters (Figure \ref{fig:entrs}). When choosing these specific forms of perturbation, we proceeded from the following principles. Firstly, an arbitrary function of interest to us from a practical point of view can be represented as a decomposition with respect to the presented elementary perturbations. Also these elementary functions allow us to consider the influence of the perturbation of the original correlation function $G_2$ in different spatial ranges, both individually and in combination. 
Thus we analyze the following perturbations:
\begin{itemize}
    \item The first perturbation is a smooth step-like function with increasing front in the horizontal coordinate $p_1$. We can treat it as $U_1(x-p_1)$.
    \item The second perturbation $U_2(\frac{x-x_0}{p_2})$ is a Gaussian function with a varying width $p_2$ but a fixed peak position and area under curve
    \item The third perturbation has a form of $F(x) = x^4e^{-\frac{x^2}{2\sigma^2}}$ and is displaced by $p_3$ to the right $U_3(x) = F(x-p_3)$
    \item  The fourth perturbation $U_4(x) = F\left(\frac{x}{p_4}\right)$ is similar to the third one but we stretch it in the horizontal direction instead of simple moving.
    
\end{itemize}   
\section{Results and Discussion}
\label{sec:res}
In the above mentioned four cases the orders of magnitude of entropy differ, since the principal components vary for different correlation functions. For this reason we normalize the entropy change by its maximum value only for the convenience of visual analysis. 

\begin{figure*}
  \centering
   \includegraphics[width = 0.95\textwidth]{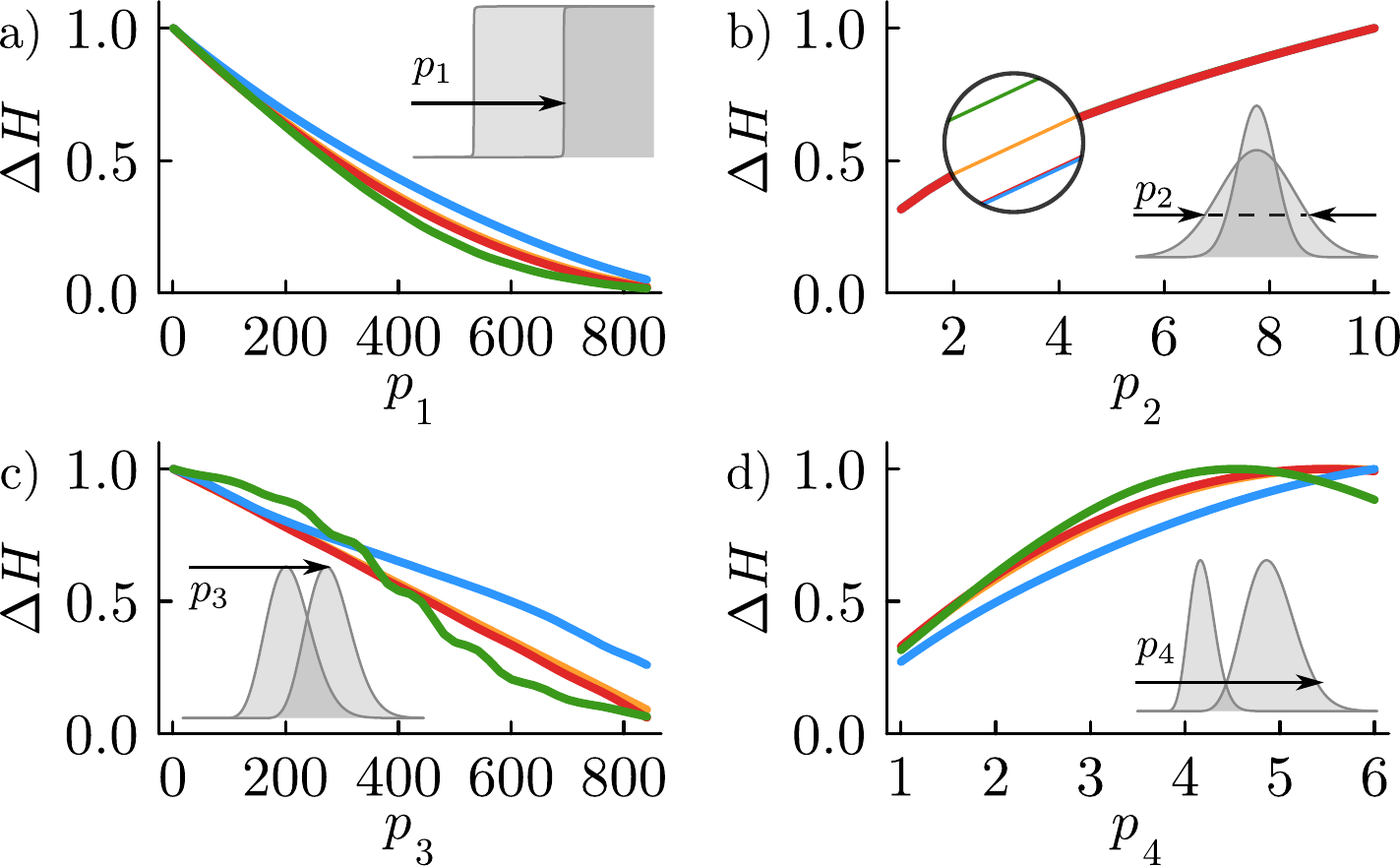}
  \caption{The entropy change with different correlation functions and different perturbations normalized by the maximum value. The gray insets show the form of perturbation: a) moving step, b) widening Gaussian, c) moving and d) widening bell-like functions. Color marks the same correlation functions as shown in figure \ref{fig:corfs}. In the case b) the plots are distinguishable only upon closer look.}
\label{fig:entrs}
\end{figure*}

We expect the entropy as a measure of information content to be higher when the perturbation has a more significant impact on the most meaningful regions in $G_2$. The initial slope of $G_2$ and the nearby surroundings appear to be such a region. The figure \ref{fig:entrs} shows the entropy change normalized by its maximum value for each of proposed perturbations. We make a few nontrivial observations from these graphs. For all types of the correlation functions the less the perturbed region of $G_2$, the less entropy is. Moreover, we see that (as expected) the information content of the different regions is not the same.

Moving the front of perturbation to the right in the sub-figure \ref{fig:entrs}.a decreases the perturbation in the region close to zero shift which results in decreasing the entropy. On the contrary, widening the perturbation in the case \ref{fig:entrs}.b increases the perturbation in the region close to zero shift and decreases the entropy change. Thus, the change of entropy increases with increasing the perturbation in the region close to zero shift. It means that, as we expected, the information content of this region is higher. The sub-figure \ref{fig:entrs}.c shows a nontrivial oscillation over existing trend for the green curve which corresponds to the period of the green $G_2$ oscillation. As for the case \ref{fig:entrs}.d we can notice that the entropy first increases and then decreases, because the perturbation, on the one hand, increases its area but, on the other hand, moves out of the region of interest. 

Thus, the analysis of the developed model on simple functions confirms our qualitative expectations of entropy for random fields. However, the main advantage of the model is its simplicity for numerical calculations. In the next section, we will try to estimate the complexity of calculating the entropy for proposed model uncertainty.

\subsection{Information Content of Heterogeneous Media Descriptors}
Our results are of particular interest in the field of porous materials research, since the statistical characterisation of heterogeneous random media is based on the use of correlation functions \cite{Torquato_book}. Two-phase random media can be characterised with an indicator function
\begin{gather}
    I(\textbf{r}) = 
    \begin{cases}
    1, & \text{if \textbf{r} is solid}\\
    0, & \text{otherwise}
    \end{cases}
\end{gather}
and the indicator function of its inter-phase boundary is defined as follows \cite{Torquato_book}:
\begin{gather}
    M(\textbf{r}) = |\nabla I(\textbf{r})|
\end{gather}
Thus the n-point function (here brackets define ensemble average over realizations):
\begin{gather}
S_n(\mathbf{r_1}\dots\mathbf{r_n})=\langle I(\mathbf{r_1})\dots I(\mathbf{r_n})\rangle
\end{gather}
has the sense of probability for certain phase of random media to exist in points $\mathbf{r_1}$ \dots $\mathbf{r_n}$ at the same time. Furthermore, surface-volume and surface-surface functions can be used to statistically characterise interposition between surface and volume and surface itself.
\begin{gather}
    F_{sv}(\mathbf{r_1}, \mathbf{r_2})=\langle M(\mathbf{r_1}) I(\mathbf{r_2})\rangle \\
    F_{ss}(\mathbf{r_1}, \mathbf{r_2})=\langle M(\mathbf{r_1}) M(\mathbf{r_2})\rangle
\end{gather}

These and other so called descriptor functions are the important tool for characterisation and stochastic reconstruction of heterogeneous porous geometry. Even more complex correlations of solid, interface and other regions can be used. For example \cite{samarin2023} suggest a robust way to calculate the n-point surface-surface correlation functions for a wide variety of cases.

Quite often characterisation should be done repeatedly as a part of an optimization process. This happens, for example, when solving inverse problem of porous media reconstruction \cite{roberts1997statistical,yeong1998reconstructing,tahmasebi2012reconstruction,karsanina2018hierarchical,cherkasov2021adaptive} with initially given statistical or any other properties. In this case execution speed of a single characterization procedure governs the execution speed of the whole algorithm. Fast Fourier Transform \cite{brigham1988fast} is a powerful tool to accelerate any correlation-based statistical descriptor which becomes a significant advantage when choosing the set of descriptors for certain problem. Thus the information content of different descriptor functions expressed in terms of Shannon entropy is a way to choose the most informative and computationally effective one for certain type of porous media structure. 
\begin{figure*}
  \centering
  \includegraphics[width = 0.95\textwidth]{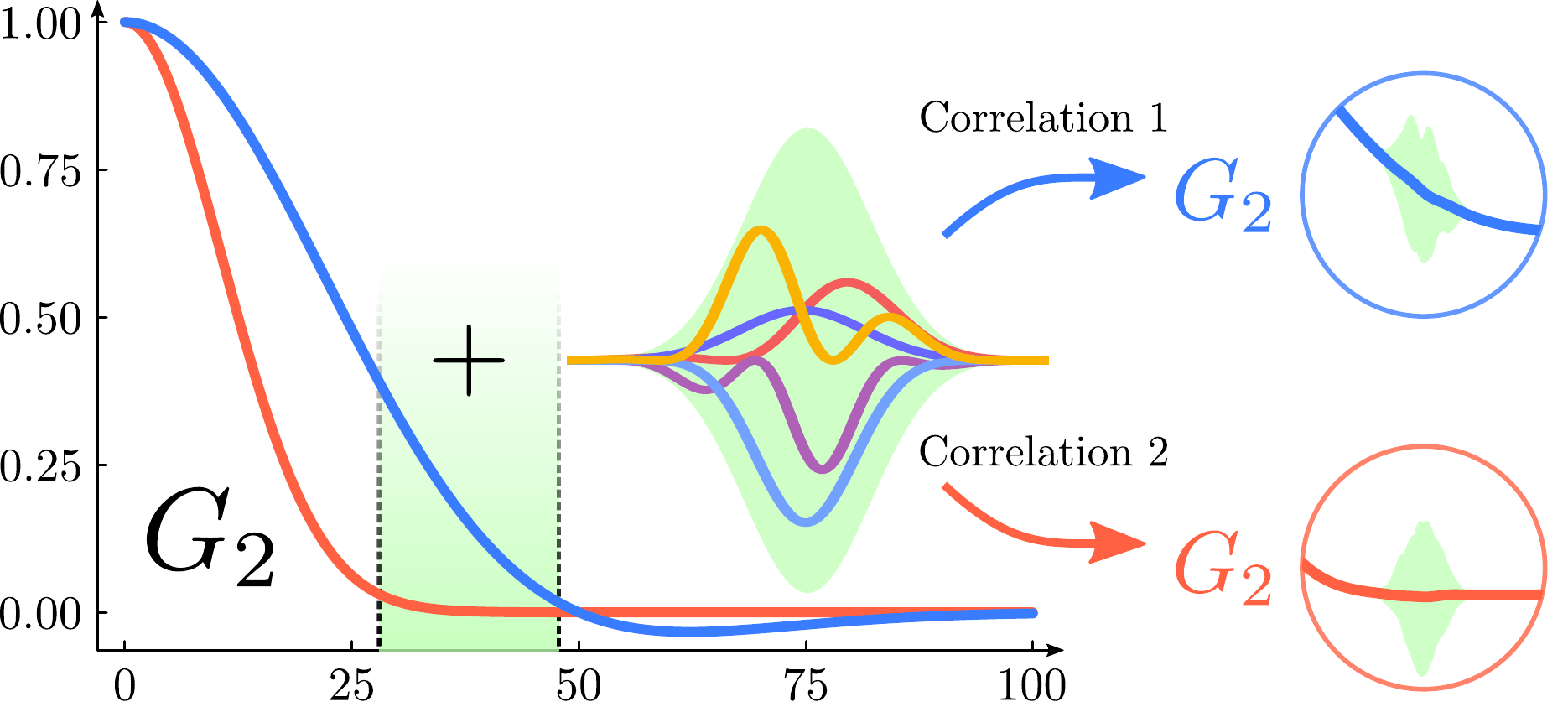}
  \caption{Perturbations from the same distribution are in different degree crucial for thick and thin correlation. Repeated sampling such perturbation together with direct reconstruction is a way to estimate the scatter of $S_2$ and $F_{ss}$ for resulting binary images}
    \label{fig:pert_sketch}
\end{figure*}
\begin{figure*}
  \centering
  \includegraphics[width = 0.95\textwidth]{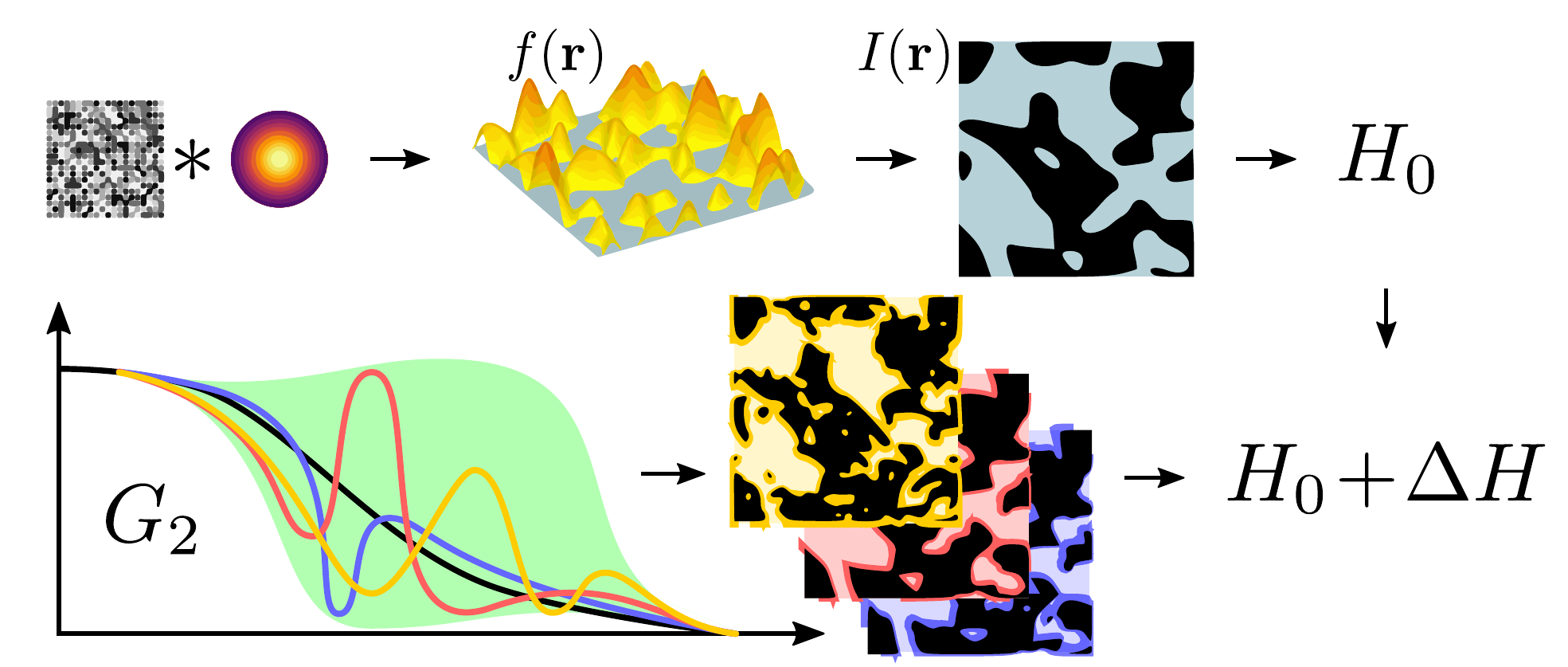}
  \caption{Random noise convolved with creating perturbation of correlation function $G_2$ produces certain realization of random field $f(\textbf{r})$. When thresholded at certain level the field becomes binary $I(\textbf{r})$. Slight perturbations of correlation function leads to slight perturbation of resulting field which results in degeneracy of field and loss of information content}
    \label{fig:pert_sketch_deg_wfl}
\end{figure*}
\begin{figure*}
  \centering
  \includegraphics[width = 0.95\textwidth]{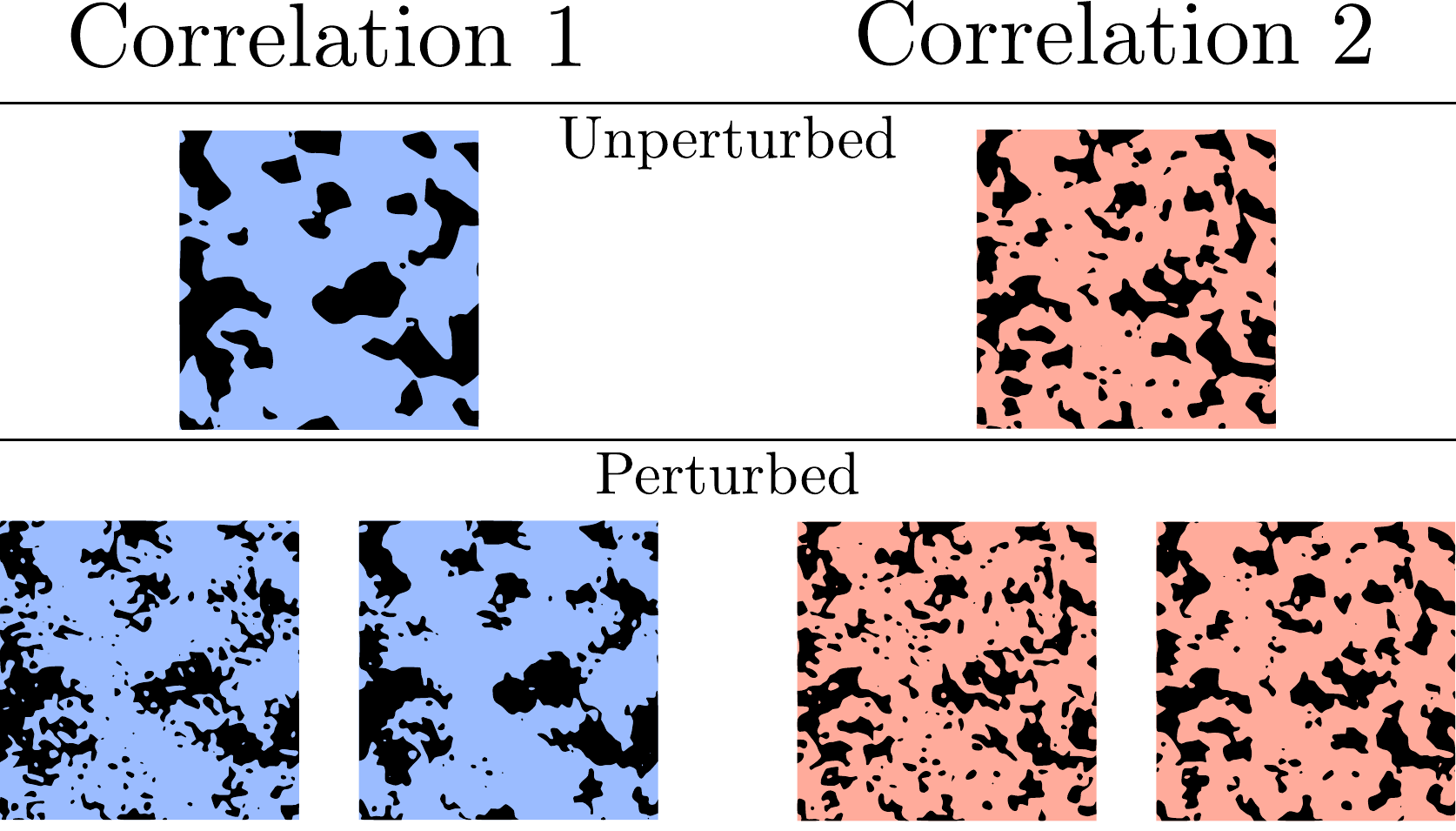}
  \caption{Results of generating binary random media with same initial noise for both thick and thin correlations 1 and 2, respectively. Top: unperturbed correlations, bottom: perturbed correlations}
    \label{fig:pert_sketch_2}
\end{figure*}
\begin{figure*}
  \centering
  \includegraphics[width = 0.95\textwidth]{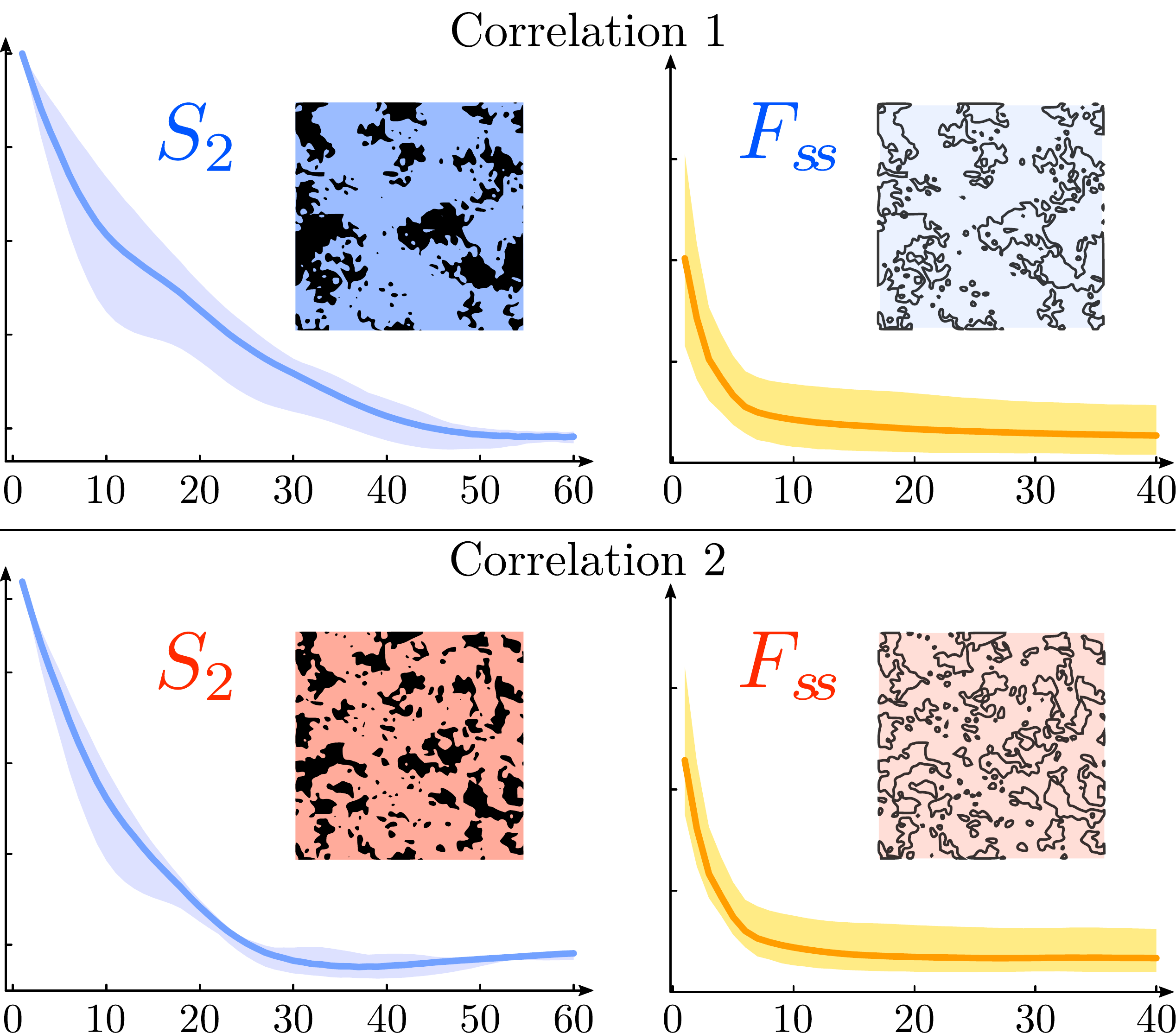}
  \caption{$S_2$ and $F_{ss}$ correlation function clouds for equally perturbed thick and thin correlations 1, 2 respectively (Figure \ref{fig:pert_sketch}). Perturbation of more meaningful regions of correlation function produces wider clouds for $S_2$ and $F_{ss}$}
    \label{fig:pert_sketch_3}
\end{figure*}

To illustrate how perturbation correlation function $G_2$ influences solid-solid $S_2$ and surface-surface $F_{ss}$ correlation functions we perform the following numerical experiment (Figure \ref{fig:pert_sketch}). For given thick  correlation function 1 which is cosine modulated by Gaussian and thin correlation function 2 which is just a decreasing Gaussian we perform a series of perturbations. Each perturbation is space and amplitude limited smooth oscillating curve (in contrast to non-correlated noise in our model). The perturbation is localised in such a region that is more meaningful for thick correlation rather than for the thin one. 
Then we use approach given in \cite{roberts1997statistical} to generate a series of reconstructions.  In particular, for each perturbed $G_2$ we obtain its spectral density and generate Gaussian random field using set of random but permanent for each perturbation spectral coefficients. The operation is fully equiavalent to convolving certain realization of white noise with kernel which correlation equals correlation of desired field (Figure \ref{fig:pert_sketch_deg_wfl}). Truncating this field at certain level produces binary structures (Figure \ref{fig:pert_sketch_2}) with slightly perturbed $S_2$ and $F_{ss}$ which now have the shape of clouds not single curves (Figure \ref{fig:pert_sketch_3}). 
When equally perturbing regions of different meaningfullness of different correlation functions the disturbed thicker correlation  produces wider clouds both for $S_2$ and $F_{ss}$. This result well coincides with the idea that perturbation of more meaningful region leads to higher impact not only on entropy but also on different descriptors. 
The developed approach allows us to estimate the information content of correlation-based descriptors and bridges the gap between information theory and heterogeneous media reconstruction. Thus, our theoretical result can be applied in further research on the topic which is of great interest.

\subsection{Computational Cost of Monte Carlo Entropy Calculations with Model Uncertainty}

Monte Carlo sampling is an alternate and commonly used way of the numerical calculation of integrals e. g. the one for entropy.
To calculate the integral of two functions product $f$ and $\rho$ (where $\rho$ defines some probability distribution) using MC one can approximate it with the finite sum of function values $f(x_i)$ at the points $x_i$ sampled from the probability density $\rho(x_i)$
\begin{gather}
 \int f(x) \rho(x) dx \approx \frac{1}{N}\sum_{i=1}^N f(x_i)
\end{gather}

The entropy can be obtained by the integral (\ref{eqn:ent})
\begin{gather}
    \label{eqn:ent_mc}
    H =  -\int dU\pi(U)\int d\chi \rho(\chi|U)\ln\left[\int dS\rho(\chi|S)\pi(S)\right]  
\end{gather}
where $\pi (U)$ is the probability density of the perturbation $U$ while $\rho(\boldsymbol{\chi}|U)$ is the conditional probability of $\boldsymbol{\chi}$ in the case of this perturbation.
To calculate this integral we need to consequently apply Monte Carlo sampling for calculation of all the internal integrals.
Their order is conditioned by the sampling convenience. The outer Monte Carlo sum
\begin{gather}
\label{eqn:ent_mc_1}
    \frac{1}{N}\sum_{i=1}^N f\left(U_i\right)
    \quad
    U_i \in \pi(U) 
\end{gather}
requires the calculation of the medium sum
\begin{gather}
\label{eqn:ent_mc_2}
    f\left(U_i\right) = \frac{1}{K}\sum_{j=1}^K g(\chi_j)
    \quad
    \chi_j \in \rho(\chi|U_i)
\end{gather}
which again requires the precalculation of
\begin{gather}
\label{eqn:ent_mc_3}
    g(\chi_j) = \ln\left[ \int dS \pi(S) \rho(\chi_j|S) \right] = 
    \ln\left[ \frac{1}{M} \sum_{k=1}^M \rho(\chi_j|S_k)\right]
    \quad
    S_k \in \pi(S)
\end{gather}

The whole algorithm seems to be extremely computationally expensive. Genuinely, even for a simple case of $5\times5$  matrix Monte Carlo summation should be done in the 10-dimensional space ($5\times(5-1)/2$ different variables) with $N\approx M\approx K\approx 10^{10}$ summands.

Moreover, the expressions (\ref{eqn:ent_mc_2}, \ref{eqn:ent_mc_3}) require $K$ and $M$ calculations of $\rho_k$ which is a complex procedure itself. Actually, singular value decomposition of the correlation matrix should be done $K\times N$ times. These estimations altogether allow us to state with complete confidence that direct Monte Carlo calculation of the entropy of Gaussian field with random correlation matrix perturbation is not so easy to obtain in reasonable time and analytical models are highly desirable. 

Monte-Carlo-based entropic approach for information content was presented in \cite{gommes2012pre}. The authors introduced energy landscape as a distance functional between certain $S_2$ functions in configuration space. They also performed Monte-Carlo sampling to obtain these landscapes for different patterns. This resulted in approximate estimate of related to entropy self-information via the energy landscape profile. This approach in terms of information content was used to characterize the degeneracy of different structures like single and hard disks, Poisson point processes and polycrystals. The analytical framework, as presented in this work, attacks the problem from a completely different angle and potentially allows to obtain information content for any 2D or 3D image without both Monte-Carlo simulation and estimations from relationships obtained with such simulations from small-scale 2D images.

%
\section{Summary and outlook}

This research paper presents a theoretical framework aimed at evaluating the information content of descriptors employed in the characterization and reconstruction of structures within random fields. Specifically, our focus lies on quantifying the information loss arising from model uncertainty during the reconstruction process. We substantiate our claim by demonstrating that stochastic perturbations introduced into the correlation function of random fields result in a decrease in their information content.

In this work, we develop a theoretical model that considers Gaussian fields possessing an arbitrary correlation matrix. Leveraging the Karhunen-Loève transform and matrix perturbation theory, we derive an analytical expression for the Shannon entropy of the perturbed random fields. Additionally, we ascertain that distinct regions of the correlation function exhibit varying levels of informativeness and sensitivity to perturbations. To exemplify the application of the analytical expression for entropy calculation, we outline a computational workflow within the paper. We explore several uncertainty models, each incorporating different parameterized perturbations, and analyze the behavior of entropy for each perturbation. This analysis reveals that the information content is more pronounced in regions where the perturbation exerts a significant impact on the correlation function.

The findings of this study bear significance on the characterization and reconstruction of heterogeneous media, particularly within the domain of porous materials research. The utilization of correlation functions as descriptors for statistical characterization is a common practice within this field. Our proposed model offers a means to estimate the information content associated with such descriptors, thereby facilitating the selection of the most informative ones.

Moreover, it is noteworthy to delineate several prospective avenues for the advancement of the developed model as well as its practical applications:
\begin{itemize}
\item Generalization to Non-Gaussian Fields:  the current model assumes Gaussian random fields. Exploring the information content and entropy analysis for non-Gaussian fields would be an interesting direction for further research. Many real-world structures exhibit non-Gaussian behavior, and understanding the impact of non-Gaussianity on the information content of descriptors could lead to more accurate characterization and reconstruction methods.
\item Application to Specific Fields: The framework presented in this research has wide applicability in various fields involving the characterization of structures. Future studies could focus on specific applications, such as material science, geophysics, or biomedical imaging. Tailoring the framework to the unique challenges and requirements of these fields would provide domain-specific insights and contribute to advancements in those areas.
\item Integration with Machine Learning Techniques: Integrating the proposed framework with machine learning techniques could lead to enhanced information content assessment and reconstruction capabilities. Exploring the use of deep learning models for automatic feature extraction and optimization of reconstruction algorithms based on the estimated information content could improve the overall performance and efficiency of the framework.
\item Real-Time Information Content Assessment: Investigating real-time information content assessment methods could be valuable in dynamic systems where structures evolve or change over time. Developing algorithms that can continuously monitor the information content of descriptors and adaptively adjust reconstruction processes based on changing structures would be a significant advancement.
\item Comparative Analysis of Descriptors: Conducting a comparative analysis of different descriptors commonly used in structure characterization could provide insights into their relative information content and effectiveness. Such an analysis would help researchers choose the most informative and suitable descriptors for specific applications, optimizing the trade-off between computational cost and accuracy.
\end{itemize}

By pursuing these research advances and improvements, the proposed framework can be further refined, validated, and applied to a wider range of real-world scenarios. Overall, this research contributes to bridging the gap between descriptor-based heterogeneous media reconstruction and information theory. The developed analytical framework offers a computationally effective way to assess the information content of descriptors applied to arbitrary structures. The theoretical results presented in this paper can be further applied in related research areas and provide valuable insights for the optimization of reconstruction algorithms in porous media and other fields.

\section{Acknowledgements}
This research was supported by the Russian Science Foundation grant 23-74-00069. Collaborative effort of the authors is within the FaT iMP (Flow and Transport in Media with Pores) research group (www.porenetwork.com). We thank Dr. Konstantin Abrosimov for suggestions and administrative work. We express sincere gratitude to our colleagues Dr. Nikolay Evstigneev and Dr. Oleg Ryabkov for fruitful discussions and comments on this work.

\bibliographystyle{elsarticle-num}
\bibliography{cas-refs}

\begin{thebibliography}{10}
\expandafter\ifx\csname url\endcsname\relax
  \def\url#1{\texttt{#1}}\fi
\expandafter\ifx\csname urlprefix\endcsname\relax\def\urlprefix{URL }\fi
\expandafter\ifx\csname href\endcsname\relax
  \def\href#1#2{#2} \def\path#1{#1}\fi

\bibitem{springel2006}
V.~Springel, C.~S. Frenk, S.~D. White, The large-scale structure of the
  universe, nature 440~(7088) (2006) 1137--1144.

\bibitem{hopkins2015new}
P.~F. Hopkins, A new class of accurate, mesh-free hydrodynamic simulation
  methods, Monthly Notices of the Royal Astronomical Society 450~(1) (2015)
  53--110.

\bibitem{balashov2021}
V.~Balashov, Dissipative spatial discretization of a phase field model of
  multiphase multicomponent isothermal fluid flow, Computers \& Mathematics
  with Applications 90 (2021) 112--124.

\bibitem{rozenbaum2014}
O.~Rozenbaum, S.~R. du~Roscoat, Representative elementary volume assessment of
  three-dimensional x-ray microtomography images of heterogeneous materials:
  Application to limestones, Physical Review E 89~(5) (2014) 053304.

\bibitem{karsanina2015}
M.~V. Karsanina, K.~M. Gerke, E.~B. Skvortsova, D.~Mallants, Universal spatial
  correlation functions for describing and reconstructing soil microstructure,
  PloS one 10~(5) (2015) e0126515.

\bibitem{ledesma2018}
R.~Ledesma-Alonso, R.~Barbosa, J.~Orteg{\'o}n, Effect of the image resolution
  on the statistical descriptors of heterogeneous media, Physical Review E
  97~(2) (2018) 023304.

\bibitem{chen2020super}
H.~Chen, X.~He, Q.~Teng, R.~E. Sheriff, J.~Feng, S.~Xiong, Super-resolution of
  real-world rock microcomputed tomography images using cycle-consistent
  generative adversarial networks, Physical Review E 101~(2) (2020) 023305.

\bibitem{prokhorov2021digital}
D.~Prokhorov, V.~Lisitsa, Y.~Bazaikin, Digital image reduction for the analysis
  of topological changes in the pore space of rock matrix, Computers and
  Geotechnics 136 (2021) 104171.

\bibitem{derossi2019}
A.~Derossi, K.~M. Gerke, M.~V. Karsanina, B.~Nicolai, P.~Verboven, C.~Severini,
  Mimicking 3d food microstructure using limited statistical information from
  2d cross-sectional image, Journal of food engineering 241 (2019) 116--126.

\bibitem{nagdalian2021}
A.~Nagdalian, I.~Rzhepakovsky, S.~Siddiqui, S.~Piskov, N.~Oboturova,
  L.~Timchenko, A.~Lodygin, A.~Blinov, S.~Ibrahim, Analysis of the content of
  mechanically separated poultry meat in sausage using computing
  microtomography, Journal of Food Composition and Analysis 100 (2021) 103918.

\bibitem{park2020}
S.~Park, S.~Lim, P.~Siriviriyakul, J.~S. Jeon, Three-dimensional pore network
  characterization of reconstructed extracellular matrix, Physical Review E
  101~(5) (2020) 052414.

\bibitem{garum2020}
M.~Garum, P.~W. Glover, P.~Lorinczi, R.~Drummond-Brydson, A.~Hassanpour,
  Micro-and nano-scale pore structure in gas shale using x$\mu$-ct and fib-sem
  techniques, Energy \& Fuels 34~(10) (2020) 12340--12353.

\bibitem{gerke2021}
K.~M. Gerke, E.~V. Korostilev, K.~A. Romanenko, M.~V. Karsanina, Going
  submicron in the precise analysis of soil structure: A fib-sem imaging study
  at nanoscale, Geoderma 383 (2021) 114739.

\bibitem{hopkins2013stars}
P.~F. Hopkins, Why do stars form in clusters? an analytic model for stellar
  correlation functions, Monthly Notices of the Royal Astronomical Society
  428~(3) (2013) 1950--1957.

\bibitem{jiao2013}
Y.~Jiao, E.~Padilla, N.~Chawla, Modeling and predicting microstructure
  evolution in lead/tin alloy via correlation functions and stochastic material
  reconstruction, Acta Materialia 61~(9) (2013) 3370--3377.

\bibitem{fomin2023soil}
D.~S. Fomin, A.~V. Yudina, K.~A. Romanenko, K.~N. Abrosimov, M.~V. Karsanina,
  K.~M. Gerke, Soil pore structure dynamics under steady-state wetting-drying
  cycle, Geoderma 432 (2023) 116401.

\bibitem{gerke2019tensor}
K.~M. Gerke, M.~V. Karsanina, R.~Katsman, Calculation of tensorial flow
  properties on pore level: Exploring the influence of boundary conditions on
  the permeability of three-dimensional stochastic reconstructions, Physical
  Review E 100~(5) (2019) 053312.

\bibitem{rozanski2023}
A.~R{\'o}{\.z}a{\'n}ski, J.~Rainer, D.~Stefaniuk, I.~Sevostianov,
  D.~{\L}yd{\.z}ba, Identification of ‘replacement’microstructure for
  porous medium from thermal conductivity measurements: Problem formulation and
  numerical solution, International Journal of Engineering Science 182 (2023)
  103788.

\bibitem{zimm1948scattering}
B.~H. Zimm, The scattering of light and the radial distribution function of
  high polymer solutions, The Journal of chemical physics 16~(12) (1948)
  1093--1099.

\bibitem{becker2010radial}
N.~Becker, A.~Rosa, R.~Everaers, The radial distribution function of worm-like
  chains, The European Physical Journal E 32 (2010) 53--69.

\bibitem{vogel2010}
H.-J. Vogel, U.~Weller, S.~Schl{\"u}ter, Quantification of soil structure based
  on minkowski functions, Computers \& Geosciences 36~(10) (2010) 1236--1245.

\bibitem{schroder2011minkowski}
G.~E. Schr{\"o}der-Turk, W.~Mickel, S.~C. Kapfer, M.~A. Klatt, F.~M. Schaller,
  M.~J. Hoffmann, N.~Kleppmann, P.~Armstrong, A.~Inayat, D.~Hug, et~al.,
  Minkowski tensor shape analysis of cellular, granular and porous structures,
  Advanced Materials 23~(22-23) (2011) 2535--2553.

\bibitem{Torquato_book}
S.~Torquato, Random heterogeneous materials: microstructure and macroscopic
  properties, Springer-Verlag New York, 2002.
\newblock \href {https://doi.org/10.1007/978-1-4757-6355-3}
  {\path{doi:10.1007/978-1-4757-6355-3}}.

\bibitem{vogel2022holistic}
H.-J. Vogel, M.~Balseiro-Romero, A.~Kravchenko, W.~Otten, V.~Pot,
  S.~Schl{\"u}ter, U.~Weller, P.~C. Baveye, A holistic perspective on soil
  architecture is needed as a key to soil functions, European Journal of Soil
  Science 73~(1) (2022) e13152.

\bibitem{yudina2023dual}
A.~Yudina, Y.~Kuzyakov, Dual nature of soil structure: The unity of aggregates
  and pores, Geoderma 434 (2023) 116478.

\bibitem{gommes2012pre}
C.~J. Gommes, Y.~Jiao, S.~Torquato, Microstructural degeneracy associated with
  a two-point correlation function and its information content, Physical Review
  E 85~(5) (2012) 051140.

\bibitem{nur1998critical}
A.~Nur, G.~Mavko, J.~Dvorkin, D.~Galmudi, Critical porosity: A key to relating
  physical properties to porosity in rocks, The Leading Edge 17~(3) (1998)
  357--362.

\bibitem{chapuis2003use}
R.~P. Chapuis, M.~Aubertin, On the use of the kozeny carman equation to predict
  the hydraulic conductivity of soils, Canadian Geotechnical Journal 40~(3)
  (2003) 616--628.

\bibitem{kozeny1927}
J.~Kozeny, Uber kapillare leitung des wassers im boden-aufstieg, versickerung
  und anwendung auf die bewasserung, sitzungsberichte der akademie der
  wissenschaften wien, Mathematisch Naturwissenschaftliche Abteilung 136 (1927)
  271--306.

\bibitem{carman1937}
P.~C. Carman, Fluid flow through a granular bed, Trans. Inst. Chem. Eng. London
  15 (1937) 150--156.

\bibitem{debye1957scattering}
P.~Debye, H.~Anderson~Jr, H.~Brumberger, Scattering by an inhomogeneous solid.
  ii. the correlation function and its application, Journal of applied Physics
  28~(6) (1957) 679--683.

\bibitem{gommes2018stochastic}
C.~J. Gommes, Stochastic models of disordered mesoporous materials for
  small-angle scattering analysis and more, Microporous and Mesoporous
  Materials 257 (2018) 62--78.

\bibitem{adler1990flow}
P.~Adler, C.~G. Jacquin, J.~Quiblier, Flow in simulated porous media,
  International Journal of Multiphase Flow 16~(4) (1990) 691--712.

\bibitem{EPL2}
K.~M. Gerke, M.~V. Karsanina, Improving stochastic reconstructions by weighting
  correlation functions in an objective function, EPL (Europhysics Letters)
  111~(5) (2015) 56002.

\bibitem{vcapek2018importance}
P.~{\v{C}}apek, On the importance of simulated annealing algorithms for
  stochastic reconstruction constrained by low-order microstructural
  descriptors, Transport in Porous Media 125~(1) (2018) 59--80.

\bibitem{tahmasebiPRL}
P.~Tahmasebi, M.~Sahimi, Cross-correlation function for accurate reconstruction
  of heterogeneous media, Physical review letters 110~(7) (2013) 078002.

\bibitem{feng2018accelerating}
J.~Feng, Q.~Teng, X.~He, X.~Wu, Accelerating multi-point statistics
  reconstruction method for porous media via deep learning, Acta Materialia 159
  (2018) 296--308.

\bibitem{gravey2020quicksampling}
M.~Gravey, G.~Mariethoz, Quicksampling v1. 0: a robust and simplified
  pixel-based multiple-point simulation approach, Geoscientific Model
  Development 13~(6) (2020) 2611--2630.

\bibitem{zhang2022improved}
F.~Zhang, Q.~Teng, X.~He, X.~Wu, X.~Dong, Improved recurrent generative model
  for reconstructing large-size porous media from two-dimensional images,
  Physical Review E 106~(2) (2022) 025310.

\bibitem{volkhonskiy2022generative}
D.~Volkhonskiy, E.~Muravleva, O.~Sudakov, D.~Orlov, E.~Burnaev, D.~Koroteev,
  B.~Belozerov, V.~Krutko, Generative adversarial networks for reconstruction
  of three-dimensional porous media from two-dimensional slices, Physical
  Review E 105~(2) (2022) 025304.

\bibitem{chubb2000every}
C.~Chubb, J.~I. Yellott, Every discrete, finite image is uniquely determined by
  its dipole histogram, Vision Research 40~(5) (2000) 485--492.

\bibitem{fullwood2008}
D.~T. Fullwood, S.~R. Niezgoda, S.~R. Kalidindi, Microstructure reconstructions
  from 2-point statistics using phase-recovery algorithms, Acta Materialia
  56~(5) (2008) 942--948.

\bibitem{dietrich1995scattering}
S.~Dietrich, A.~Haase, Scattering of x-rays and neutrons at interfaces, Physics
  Reports 260~(1-2) (1995) 1--138.

\bibitem{li2018direct}
H.~Li, S.~Singh, N.~Chawla, Y.~Jiao, Direct extraction of spatial correlation
  functions from limited x-ray tomography data for microstructural
  quantification, Materials Characterization 140 (2018) 265--274.

\bibitem{karsanina2021compressing}
M.~V. Karsanina, E.~V. Lavrukhin, D.~S. Fomin, A.~V. Yudina, K.~N. Abrosimov,
  K.~M. Gerke, Compressing soil structural information into parameterized
  correlation functions, European Journal of Soil Science 72~(2) (2021)
  561--577.

\bibitem{jiao2014modeling}
Y.~Jiao, N.~Chawla, Modeling and characterizing anisotropic inclusion
  orientation in heterogeneous material via directional cluster functions and
  stochastic microstructure reconstruction, Journal of Applied Physics 115~(9)
  (2014) 093511.

\bibitem{gerke2014improving}
K.~M. Gerke, M.~V. Karsanina, R.~V. Vasilyev, D.~Mallants, Improving pattern
  reconstruction using directional correlation functions, Europhysics Letters
  106~(6) (2014) 66002.

\bibitem{yao1993high}
J.~Yao, P.~Frykman, F.~Kalaydjian, J.~Thovert, P.~Adler, High-order moments of
  the phase function for real and reconstructed model porous media: a
  comparison, Journal of colloid and interface science 156~(2) (1993) 478--490.

\bibitem{vcapek2011transport}
P.~{\v{C}}apek, V.~Hejtm{\'a}nek, J.~Kolafa, L.~Brabec, Transport properties of
  stochastically reconstructed porous media with improved pore connectivity,
  Transport in porous media 88 (2011) 87--106.

\bibitem{karsanina2018enhancing}
M.~V. Karsanina, K.~M. Gerke, E.~B. Skvortsova, A.~L. Ivanov, D.~Mallants,
  Enhancing image resolution of soils by stochastic multiscale image fusion,
  Geoderma 314 (2018) 138--145.

\bibitem{adam2022efficient}
A.~Adam, F.~Wang, X.~Li, Efficient reconstruction and validation of
  heterogeneous microstructures for energy applications, International Journal
  of Energy Research 46~(15) (2022) 22757--22771.

\bibitem{jiao2009superior}
Y.~Jiao, F.~Stillinger, S.~Torquato, A superior descriptor of random textures
  and its predictive capacity, Proceedings of the National Academy of Sciences
  106~(42) (2009) 17634--17639.

\bibitem{campaigne2012frozen}
W.~R. Campaigne, P.~W. Fieguth, Frozen-state hierarchical annealing, IEEE
  transactions on image processing 22~(4) (2012) 1486--1497.

\bibitem{karsanina2018hierarchical}
M.~V. Karsanina, K.~M. Gerke, Hierarchical optimization: Fast and robust
  multiscale stochastic reconstructions with rescaled correlation functions,
  Physical review letters 121~(26) (2018) 265501.

\bibitem{wang2001efficient}
F.~Wang, D.~P. Landau, Efficient, multiple-range random walk algorithm to
  calculate the density of states, Physical review letters 86~(10) (2001) 2050.

\bibitem{gommes2012prl}
C.~J. Gommes, Y.~Jiao, S.~Torquato, Density of states for a specified
  correlation function and the energy landscape, Physical review letters
  108~(8) (2012) 080601.

\bibitem{chen2020probing}
P.-E. Chen, W.~Xu, Y.~Ren, Y.~Jiao, Probing information content of hierarchical
  n-point polytope functions for quantifying and reconstructing disordered
  systems, Physical Review E 102~(1) (2020) 013305.

\bibitem{skolnick2021understanding}
M.~Skolnick, S.~Torquato, Understanding degeneracy of two-point correlation
  functions via debye random media, Physical Review E 104~(4) (2021) 045306.

\bibitem{thovert2020influence}
J.-F. Thovert, V.~V. Mourzenko, On the influence of boundary conditions when
  determining transport coefficients from digital images of heterogeneous
  media., Advances in Water Resources 141 (2020) 103612.

\bibitem{scandelli2022computation}
H.~Scandelli, A.~Ahmadi-Senichault, C.~Levet, J.~Lachaud, Computation of the
  permeability tensor of non-periodic anisotropic porous media from 3d images,
  Transport in Porous Media 142~(3) (2022) 669--697.

\bibitem{chen2022impacts}
H.~Chen, X.~Wu, M.~Han, Y.~Zhang, Impacts of solid wall boundary conditions in
  the lattice boltzmann method on turbulent outdoor flow: A case study of a
  single 1: 1: 2 building model, Building and Environment 226 (2022) 109708.

\bibitem{zhang2000pore}
D.~Zhang, R.~Zhang, S.~Chen, W.~E. Soll, Pore scale study of flow in porous
  media: Scale dependency, rev, and statistical rev, Geophysical research
  letters 27~(8) (2000) 1195--1198.

\bibitem{gerke2021pore}
K.~M. Gerke, M.~V. Karsanina, How pore structure non-stationarity compromises
  flow properties representativity (rev) for soil samples: Pore-scale modelling
  and stationarity analysis, European Journal of Soil Science 72~(2) (2021)
  527--545.

\bibitem{ghanbarian2022estimating}
B.~Ghanbarian, Estimating the scale dependence of permeability at pore and core
  scales: Incorporating effects of porosity and finite size, Advances in Water
  Resources 161 (2022) 104123.

\bibitem{tahmasebi2015}
P.~Tahmasebi, M.~Sahimi, Reconstruction of nonstationary disordered materials
  and media: Watershed transform and cross-correlation function, Physical
  Review E 91~(3) (2015) 032401.

\bibitem{gommes2009}
C.~J. Gommes, J.-P. Pirard, Morphological models of complex ordered materials
  based on inhomogeneously clipped gaussian fields, Physical Review E 80~(6)
  (2009) 061401.

\bibitem{karsanina2023stochastic}
M.~V. Karsanina, K.~M. Gerke, Stochastic (re) constructions of non-stationary
  material structures: Using ensemble averaged correlation functions and
  non-uniform phase distributions, Physica A: Statistical Mechanics and its
  Applications 611 (2023) 128417.

\bibitem{SciRep}
K.~M. Gerke, M.~V. Karsanina, D.~Mallants, Universal stochastic multiscale
  image fusion: an example application for shale rock, Scientific reports 5~(1)
  (2015) 1--12.

\bibitem{Havelka}
J.~Havelka, A.~Ku{\v{c}}erov{\'a}, J.~S{\`y}kora, Compression and
  reconstruction of random microstructures using accelerated lineal path
  function, Computational Materials Science 122 (2016) 102--117.

\bibitem{kamrava2020linking}
S.~Kamrava, P.~Tahmasebi, M.~Sahimi, Linking morphology of porous media to
  their macroscopic permeability by deep learning, Transport in Porous Media
  131~(2) (2020) 427--448.

\bibitem{roding2020predicting}
M.~R{\"o}ding, Z.~Ma, S.~Torquato, Predicting permeability via statistical
  learning on higher-order microstructural information, Scientific reports
  10~(1) (2020) 1--17.

\bibitem{cheng2022data}
S.~Cheng, Y.~Jiao, Y.~Ren, Data-driven learning of 3-point correlation
  functions as microstructure representations, Acta Materialia 229 (2022)
  117800.

\bibitem{EfimEPL}
E.~Lavrukhin, M.~Karsanina, K.~Gerke, Stochastic reconstruction of particulate
  media using simulated annealing: improving pore connectivity, EPL (2023).

\bibitem{shannon1948mathematical}
C.~E. Shannon, A mathematical theory of communication, The Bell system
  technical journal 27~(3) (1948) 379--423.

\bibitem{hristopulos2020random}
D.~T. Hristopulos, Random fields for spatial data modeling, Springer, 2020.

\bibitem{ma2011principal}
S.~Ma, Y.~Dai, Principal component analysis based methods in bioinformatics
  studies, Briefings in bioinformatics 12~(6) (2011) 714--722.

\bibitem{stacklies2007pcamethods}
W.~Stacklies, H.~Redestig, M.~Scholz, D.~Walther, J.~Selbig, pcamethods—a
  bioconductor package providing pca methods for incomplete data,
  Bioinformatics 23~(9) (2007) 1164--1167.

\bibitem{mishra2017multivariate}
S.~P. Mishra, U.~Sarkar, S.~Taraphder, S.~Datta, D.~Swain, R.~Saikhom,
  S.~Panda, M.~Laishram, Multivariate statistical data analysis-principal
  component analysis (pca), International Journal of Livestock Research 7~(5)
  (2017) 60--78.

\bibitem{brunton2016discovering}
S.~L. Brunton, J.~L. Proctor, J.~N. Kutz, Discovering governing equations from
  data by sparse identification of nonlinear dynamical systems, Proceedings of
  the national academy of sciences 113~(15) (2016) 3932--3937.

\bibitem{carlberg2018conservative}
K.~Carlberg, Y.~Choi, S.~Sargsyan, Conservative model reduction for
  finite-volume models, Journal of Computational Physics 371 (2018) 280--314.

\bibitem{sarma2006efficient}
P.~Sarma, L.~J. Durlofsky, K.~Aziz, W.~H. Chen, Efficient real-time reservoir
  management using adjoint-based optimal control and model updating,
  Computational Geosciences 10~(1) (2006) 3.

\bibitem{elizarev2021objective}
M.~Elizarev, A.~Mukhin, A.~Khlyupin, Objective-sensitive principal component
  analysis for high-dimensional inverse problems, Computational Geosciences 25
  (2021) 2019--2031.

\bibitem{vladimirov2019quantum}
I.~G. Vladimirov, I.~R. Petersen, M.~R. James, A quantum karhunen-loeve
  expansion and quadratic-exponential functionals for linear quantum stochastic
  systems, in: 2019 IEEE 58th Conference on Decision and Control (CDC), IEEE,
  2019, pp. 425--430.

\bibitem{hirschfelder1964recent}
J.~O. Hirschfelder, W.~B. Brown, S.~T. Epstein, Recent developments in
  perturbation theory, Advances in quantum chemistry 1 (1964) 255--374.

\bibitem{landau2013quantum}
L.~D. Landau, E.~M. Lifshitz, Quantum mechanics: non-relativistic theory,
  Vol.~3, Elsevier, 2013.

\bibitem{golub2013matrix}
G.~H. Golub, C.~F. Van~Loan, Matrix computations, JHU press, 2013.

\bibitem{samarin2023}
A.~Samarin, V.~Postnicov, M.~V. Karsanina, E.~V. Lavrukhin, D.~Gafurova, N.~M.
  Evstigneev, A.~Khlyupin, K.~M. Gerke, Robust surface-correlation-function
  evaluation from experimental discrete digital images, Phys. Rev. E [accepted]
  (2023).

\bibitem{roberts1997statistical}
A.~P. Roberts, Statistical reconstruction of three-dimensional porous media
  from two-dimensional images, Physical Review E 56~(3) (1997) 3203.

\bibitem{yeong1998reconstructing}
C.~Yeong, S.~Torquato, Reconstructing random media, Physical review E 57~(1)
  (1998) 495.

\bibitem{tahmasebi2012reconstruction}
P.~Tahmasebi, M.~Sahimi, Reconstruction of three-dimensional porous media using
  a single thin section, Physical Review E 85~(6) (2012) 066709.

\bibitem{cherkasov2021adaptive}
A.~Cherkasov, A.~Ananev, M.~Karsanina, A.~Khlyupin, K.~Gerke, Adaptive
  phase-retrieval stochastic reconstruction with correlation functions:
  Three-dimensional images from two-dimensional cuts, Physical Review E 104~(3)
  (2021) 035304.

\bibitem{brigham1988fast}
E.~O. Brigham, The fast Fourier transform and its applications, Prentice-Hall,
  Inc., 1988.

\bibitem{cramer1999mathematical}
H.~Cram{\'e}r, Mathematical methods of statistics, Vol.~26, Princeton
  university press, 1999.

\end{thebibliography}
\appendix

\section{Perturbed Probability Density}
\label{ap:1}
In this section, we evaluate a simple expression for the perturbed probability density of $\chi_k$. We use Taylor series expansion up to the second order of infinitesimals
\begin{gather}
    \frac{1}{\sqrt{1+x}}\cdot e^{-\frac{\alpha}{1+x}} = \frac{1}{e}\left(1 +\frac{(2\alpha - 1)x}{2} + \frac{(4\alpha^2 -12\alpha + 3)x^2}{8}\right)
\end{gather}

Here we made the following variable change $a = -\frac{\chi_k^2}{2\lambda_{k}}$ and now equation (\ref{eq:fuldens}) becomes
\begin{gather}
\rho_k\left(\chi_k|\hat{U}\right) = 
\rho_k\left(\chi_i, \lambda_k + \lambda_{k,1} + \lambda_{k,2}\right) =
\\
= \frac{1}{\sqrt{2\pi(\lambda_{k} + \lambda_{k,1} + \lambda_{k,2})} }\exp\left[-\frac{\chi_k^2}{2(\lambda_{k} + \lambda_{k,1} + \lambda_{k,2})}\right] = \nonumber\\
=\frac{e^a}{\sqrt{2\pi\lambda_{k}}} \cdot \left(1 - \left(\frac{1}{2} + a\right) \frac{\lambda_{k,1}}{\lambda_{k}}
-\left(a + \frac{1}{2}\right)\frac{\lambda_{k,2}}{\lambda_{k}} + \left(\frac{3}{8} + \frac{3}{2}a +\frac{a^2}{2}\right)\frac{\lambda_{k,1}}{\lambda_{k}}^2\right) = \nonumber\\
=\rho_k(\chi_k, \lambda_k)\left(1 + {f}_k(\chi_k, \hat{U})\right)\nonumber
\end{gather}
where the coefficients in the introduced correction ${f_k} = A(\hat{U}) + B(\hat{U})\chi_k^2 + C(\hat{U})\chi_k^4$ have the following form
\begin{gather}
A = \frac{1}{\lambda_k}\left( - \frac{1}{2}\lambda_{k,1}- \frac{1}{2}\lambda_{k,2}+  \frac{3}{8}\lambda_{k,1}^2 \right)
\label{eq:ax}
\\ \nonumber
B = \frac{1}{2\lambda_{k}^2}\left({\lambda_{k,1}} + {\lambda_{k,2}} -  \frac{3}{2}{\lambda^2_{k,1}}\right)\\
C = \frac{1}{8\lambda_{k}^3}{\lambda^2_{k,1}} \nonumber
\end{gather}

\section{Perturbed Entropy}
\label{ap:2}
In this section, we provide a detailed evaluation of the desired entropy.
First, we will show that $H_2$ does not contribute to the entropy expression
\begin{gather}
H_2 = - \int d\boldsymbol{\chi} \sum_{i\neq j}\langle{f_i}{f_j}\rangle_{\hat{U}}\prod_k \rho_{k} - 
\int d \boldsymbol{\chi} \sum_i\ln(\rho_{i})\sum_{i\neq j}\langle{f_i}{f_j}\rangle_{\hat{U}}\prod_k \rho_{k}
\end{gather}

Only the product of first-order corrections in \ref{eq:ax} leads to the expansion of $ \langle f_i f_j \rangle$ up to the second order
\begin{gather}
    \langle f_i f_j \rangle_{\hat{U}} = \frac{\langle \lambda_{i,1}, \lambda_{j,1} \rangle_{\hat{U}}}{\lambda_i, \lambda_j}\left(1-\frac{\chi_i^2}{\lambda_i}\right)\left(1-\frac{\chi_j^2}{\lambda_j}\right)
\end{gather}

The first term of $H_2$ decomposes to the sum of zero integrals product
\begin{gather}
    \int d\boldsymbol{\chi} \sum_{i\neq j}\frac{\langle \lambda_{i,1}, \lambda_{j,1} \rangle_{\hat{U}}}{\lambda_i, \lambda_j}\left(1-\frac{\chi_i^2}{\lambda_i}\right)\left(1-\frac{\chi_j^2}{\lambda_j}\right)\prod_k \rho_{k} =\\
    =  \sum_{i \neq j}\frac{\langle \lambda_{i,1}, \lambda_{j,1} \rangle_{\hat{U}}}{\lambda_i, \lambda_j} \int d\chi_i \rho_i \left(1-\frac{\chi_i^2}{\lambda_i}\right) \int d\chi_j \rho_j \left(1-\frac{\chi_j^2}{\lambda_j}\right) = \\
    =\sum_{i \neq j}\frac{\langle \lambda_{i,1}, \lambda_{j,1} \rangle_{\hat{U}}}{\lambda_i, \lambda_j} (1-1)(1-1)=0 \nonumber
\end{gather}

The second term is also the sum of vanishing products
\begin{gather}
    \int d \boldsymbol{\chi} \sum_l\ln(\rho_{l})\sum_{i\neq j}\langle{f_i}{f_j}\rangle_{\hat{U}}\prod_k \rho_{k} =\\
    = \nonumber \sum_{i \neq j, l}\mathds{E}_{\chi_l}\ln(\rho_l)\frac{\langle \lambda_{i,1}, \lambda_{j,1} \rangle_{\hat{U}}}{\lambda_i, \lambda_j} \int d\chi_i \rho_i\left(1-\frac{\chi_i^2}{\lambda_i}\right)\int d\chi_j \rho_j \left(1-\frac{\chi_j^2}{\lambda_j}\right) +\\\nonumber
    +\sum_{i \neq j}\frac{\langle \lambda_{i,1}, \lambda_{j,1} \rangle_{\hat{U}}}{\lambda_i, \lambda_j} \int d\chi_i \rho_i \ln(\rho_i) \left(1-\frac{\chi_i^2}{\lambda_i}\right)\int d\chi_j \rho_j \left(1-\frac{\chi_j^2}{\lambda_j}\right) = \\ \nonumber
    =\sum_{i \neq j}\mathds{E}_{\chi_k}\ln(\rho_l)\frac{\langle \lambda_{i,1}, \lambda_{j,1} \rangle_{\hat{U}}}{\lambda_i, \lambda_j}\left(1-1\right) \left(1-1\right) +\\\nonumber
    + \sum_{i \neq j}\frac{\langle \lambda_{i,1}, \lambda_{j,1} \rangle_{\hat{U}}}{\lambda_i, \lambda_j} \int d\chi_i \rho_i \ln(\rho_i) \left(1-\frac{\chi_i^2}{\lambda_i}\right)(1-1)=0
\end{gather}

Thus, the entropy expression (\ref{eqn:ent_raw}) contains the following terms
\begin{gather}
H = -\int d \boldsymbol{\chi} \left(1 + \sum_i \tilde{f_i}\right)\left(\sum_j\ln(\rho_{j}) + \sum_j \tilde{f_j}\right)\prod_k\rho_{k} = \\
=-\int d \boldsymbol{\chi} \sum_i\ln(\rho_{i})\prod_k \rho_{k} - \int d\boldsymbol{\chi}\sum_i \tilde{f}_{i}\prod_k \rho_{k} - 
\int d \boldsymbol{\chi} \sum_i\ln(\rho_{i})\sum_j \nonumber \tilde{f_j}\prod_k \rho_{k} 
\end{gather}

One may integrate the second term in a way similar to \ref{eqn:entr_unp} leading to
\begin{gather}
\int d\boldsymbol{\chi}  \sum_i \tilde{f}_{i}\prod_k \rho_{k}=\sum_i \int d\chi_i \rho_{i} \tilde{f}_{i}
\label{eqn:entr_sec}
\end{gather}

We also write the third term as a sum of two terms. The first one is a product which contains incomplete sum for unperturbed entropy $H_0$ except the $j$-th term. The second term is also a product with the $j$-th term itself inside it
\begin{gather}
-\int d \boldsymbol{\chi} \sum_i\ln(\rho_{i})\sum_j \tilde{f_j} \prod_k \rho_{k} =\\
= -\int d \boldsymbol{\chi}  \sum_{i \neq j} \ln(\rho_{i})\sum_{j} \tilde{f_j}\prod_k \rho_{k} - \int d \boldsymbol{\chi} \sum_j\ln(\rho_{j})\tilde{f_j}\prod_k \rho_{k} = \nonumber \\
= \left(H_0+\int d\chi_j \rho_{j} \ln(\rho_{j})\right) \sum_{j}\int d\chi_j \rho_{j} \tilde{f}_{j} -\sum_j\int d \chi_j \rho_{j}\ln(\rho_{j})\tilde{f_j} \nonumber
\label{eqn:entr_thrd}
\end{gather}

Considering the above-mentioned simplifications we obtain the intermediate equation for the perturbed entropy
\begin{gather}
H=H_0 - \sum d\chi_i \rho_{i}\tilde{f_i} + \sum_i \int d\chi_i \rho_{i}\tilde{f}_i\left(H_0 + \int d\chi_i\rho_{i}\ln(\rho_{i})\right) - \sum_i\int d\chi_i \rho_{i}\ln(\rho_{i})\tilde{f}_{i} =\nonumber\\
= H_0 + \sum_i \mathds{E}_{\chi_i}\left( \tilde{f_i}\left[H_0 - 1 +\mathds{E}_{\chi_i}\ln(\rho_{i})\right]\right) - \sum_i\mathds{E}_{\chi_i}\left[\ln(\rho_{i}) \tilde{f_i}\right]
\label{eqn:ent_tobeav}
\end{gather}

The final result is the function of only the averaged coefficients $\langle A \rangle_{\hat{U}}$, $\langle B \rangle_{\hat{U}}$ and $\langle C \rangle_{\hat{U}}$, which themselves are the functions of $ \langle\lambda_{k,1}\rangle_{\hat{U}}$, $\langle\lambda_{k,2}\rangle_{\hat{U}}$, $\langle\lambda_{k,1}^2\rangle_{\hat{U}}$
\begin{gather}
H =H_0 + \sum_i \left(H_0 - 1 +\int d\chi_i \rho_{i} \ln(\rho_{i})\right) \int d\chi_i \rho_i \left(\langle A \rangle_{\hat{U}} + \langle B \rangle_{\hat{U}} \chi_i^2 + \langle C \rangle_{\hat{U}} \chi_i^4 \right) - \nonumber \\
- \sum_i \int d\chi_i \ln(\rho_{i})\left(\langle A \rangle_{\hat{U}} + \langle B \rangle_{\hat{U}} \chi_i^2 + \langle C \rangle_{\hat{U}} \chi_i^4 \right)\nonumber 
\end{gather}

To find the expected value of terms in (\ref{eqn:ent_tobeav}) one may use several common integrals \cite{cramer1999mathematical}
\begin{gather}
\mathds{E}_{\chi_i} \ln(\rho_{i}) = \int\frac{1}{\sqrt{2\pi\lambda_{i}}}e^{-\frac{\chi_i^2}{2\lambda_{i}}}(-\frac{\chi_i^2}{2\lambda_{i}} -\ln\sqrt{2\pi\lambda_{i}})d\chi_i = \frac{1}{2}(-\ln(2\pi\lambda_{i}) - 1)\\ \nonumber
\mathds{E}_{\chi_i}\left(\ln(\rho_{i})\chi_i^2\right) = -\frac{3}{2}\lambda_{i}-\frac{1}{2}\ln(2\pi\lambda_{i})\lambda_{i}\\ \nonumber
\mathds{E}_{\chi_i}\left(\ln(\rho_{i})\chi_i^4\right) = -\frac{15}{2}\lambda_{i}^2-\frac{3}{2}\ln(2\pi\lambda_{i})\lambda_{i}^2\\ \nonumber
\mathds{E}_{\chi_i}\left(\chi_i^2 \right) =\lambda_i\\ \nonumber
\mathds{E}_{\chi_i}\left(\chi_i^4 \right) =3\lambda_i^2
\end{gather}

Substituting of common integrals into the expression for entropy we obtain
\begin{gather}
H=H_0 + \sum_k \Bigg[ \left(- \frac{1}{2}\frac{\lambda_{k,1}}{\lambda_{k}} - \frac{1}{2}\frac{\lambda_{k,2}}{\lambda_{k}} +  \frac{3}{8}\frac{\lambda_{k,1}^2}{\lambda_{k}^2}\right)\frac{1}{2}\left(-\ln(2\pi\lambda_{k}) - 1\right) +\\\nonumber
+\left(\frac{1}{2}\frac{\lambda_{k,1}}{\lambda_{k}^2} +\frac{1}{2} \frac{\lambda_{k,2}}{\lambda_{k}^2} -  \frac{3}{4}\frac{\lambda_{k,1}^2}{\lambda_{k}^3}\right)\left(-\frac{3}{2}\lambda_{k}-\frac{1}{2}\ln(2\pi\lambda_{k})\lambda_{k}\right) +\\\nonumber
+ \frac{1}{8}\frac{\lambda_{k,1}^2}{\lambda_{k}^4}\left(-\frac{15}{2}\lambda_{k}^2-\frac{3}{2}\ln(2\pi\lambda_{k})\lambda_{k}^2\right) \Bigg]
\end{gather}

Simplifying the above equation we obtain the following final expression for entropy, which depends on the averaged eigenvalues corrections
\begin{gather}
H=H_0 +\frac{1}{2}\sum_k \left(\frac{\langle\lambda_{k,1}\rangle_{\hat{U}}}{\lambda_{k}} + \frac{\langle\lambda_{k,2}\rangle_{\hat{U}}}{\lambda_{k}}+0\cdot\frac{\langle\lambda_{k,1}^2\rangle_{\hat{U}}}{\lambda_{k}}\right)
\label{eqn:ent_so}
\end{gather}

\section{Averaging Eigenvalue Corrections}
In this section, we provide the exact expressions for the averaged eigenvalue corrections $\langle\lambda_{k,1}\rangle_{\hat{U}}$ and $\langle\lambda_{k,2}\rangle_{\hat{U}}$ over realizations of $\hat{U}$. While averaging of $\lambda_{k,1}$ and taking into account the zero expectation $\mathds{E}(U(x)) = 0$ one may easily obtain
\begin{gather}
\langle\lambda_{k,1}\rangle_{\hat{U}}=\langle P_{kk}\rangle_{\hat{U}}=0
\end{gather}
where the matrix element of $\hat{U}$ is defined in the following way
\begin{gather}
P_{km} = (\psi^k, \hat{U} \psi^m )=\sum_{ij}\psi^k_i\hat{U}_{ij} \psi^m_j
\end{gather}

Now let us consider the second-order correction to the eigenvalue
\begin{gather}
\lambda_{k,2} = \sum_{m \neq k}\frac{P_{km} P_{mk}}{\lambda_k - \lambda_m}
\end{gather}

According to our model of delta correlated values of random $U(x)$ in different points in space $\langle U(x) U(x+\tau)\rangle_{\hat{U}}=u(x)\delta(\tau)$ we suggest $\langle\hat{U}_{km}\hat{U}_{ln}\rangle$ is non-zero and equal to ${u}_{ln}$ if and only if $\hat{U}_{km}$ is the same as $\hat{U}_{ln}$ or $\hat{U}_{nl}$  which can be written as follows
\begin{gather}
\langle\hat{U}_{km}\hat{U}_{ln}\rangle = (\delta_{kl}\delta_{mn} + \delta_{kn}\delta_{ml})u_{ln}
\end{gather}

Using this correlation model on entries of random matrix $\hat{U}$ we can obtain the following expression for the product of matrix elements in $\lambda_{k,2}$ 
\begin{gather}
\langle P_{k,m}P_{m,k}\rangle = 
\sum_{l,p}\left\langle\psi^k_l\sum_n\hat{U}_{l,n}\psi^m_n\psi^m_p\sum_s\hat{U}_{p,s}\psi^k_s\right\rangle_{\hat{U}}\\ \nonumber
=\sum_{l,n,p,s}\psi^k_l\psi^m_p\psi^m_n\psi^k_s(\delta_{lp}\delta_{ns}u_{ln} + \delta_{ls}\delta_{np}u_{ln}) \\ \nonumber=\sum_{l, n}\psi^k_l\psi^m_lu_{ln}\psi^k_n\psi^m_n + \sum_{l, n}(\psi^k_l)^2u_{ln}(\psi^m_n)^2
\end{gather}

Substitution of this result to the averaged second-order correction $\lambda_{k,2}$ leads to
\begin{gather}
\sum_k \frac{\langle\lambda_{k,2}\rangle}{\lambda_{k}} =  \sum_k \frac{1}{\lambda_{k}}\sum_m\frac{\langle P_{km} P_{mk}\rangle}{\lambda_{k} - \lambda_{m}}\\ =  \sum_{k,m} \frac{1}{\lambda_{k}(\lambda_k - \lambda_m)}\sum_{l,n}\left(\psi^k_l\psi^m_lu_{ln}\psi^k_n\psi^m_n+(\psi^k_l)^2u_{ln}(\psi^m_n)^2\right)
\label{eq:scp}
\end{gather}

Thus, the final expression for entropy becomes
\begin{gather}
H = H_0 +  \sum_{k,m} \frac{1}{2\lambda_{k}(\lambda_k - \lambda_m)}\sum_{l,n}\left(\psi^k_l\psi^m_lu_{ln}\psi^k_n\psi^m_n+(\psi^k_l)^2u_{ln}(\psi^m_n)^2\right)
\label{eqn:ent_sfin_2}
\end{gather}
 
In the next section we'll provide some simplification of this expression.

\section{Approximate Expression for the Entropy}
Firstly it can be seen that $\hat{U}$ is symmetric matrix with non negative entries: $U_{ij}\geq 0$ and $U_{ij} = u(|i - j|)$. This means $\hat{U}$ can be presented in the following way
\label{ineq_cor}
\begin{gather}
\hat{U} = \sum_r u_{r}\hat{D_r}\\
\hat{D_r} = 
\begin{pmatrix}
0 & \  & 1 & \  & 0\\
\  & \ddots & \  & \ddots & \  \\
1 & \  & 0 & \  & 1 \\
\  & \ddots & \  & \ddots & \  \\
0 & \  & 1 & \  & 0
\end{pmatrix}
\end{gather}
where we introduced matrix $D_r$ with nonzero entries on $r$-diagonals. Thus, one may consider scalar products in the right hand side of \ref{eq:scp}. We denote the first and the second terms in the r.h.s. by $I_1$ and $I_2$ 
\begin{gather}
I_1 = 2\sum_r\sum_{i=0}^{N-r}\psi_{1+i}^k\psi_{1+i}^m u_r \psi_{r+i}^k\psi_{r+i}^m = 2\sum_ru_r\sum_{i=0}^{N-r}x_iy_i\label{eq:sc1}\\
I_2 = \sum_r\sum_{i=0}^{N-r}\psi_{1+i}^m\psi_{1+i}^m u_r \psi_{r+i}^k\psi_{r+i}^k + \sum_r\sum_{i=0}^{N-r}\psi_{1+i}^k\psi_{1+i}^k u_r \psi_{r+i}^m\psi_{r+i}^m\label{eq:sc2}\\
=\sum_ru_r\sum_{i=0}^{N-r}(x_i^2 + y_i^2)\nonumber
\end{gather}

Here we use the following notation $x_i = \psi_{1+i}^k\psi_{r+i}^m$ and $y_i = \psi_{1+i}^m\psi_{r+i}^k$. Next, with a help of the following simple inequality
\begin{gather}
    2|x_iu_i|\leq|x_i^2+y_i^2|=x_i^2+y_i^2
\end{gather}
it is easy to compare \ref{eq:sc1} and \ref{eq:sc2}: $|I_1| < |I_2|$.
Moreover, taking the constant values in matrix $U_{ij}=u$ results in $I_1=0$ and $I_2=1$ due to normality condition $(\psi^i \psi^j)=\delta_{ij}$. 

These two arguments together with variety of numerical comparisons allow us to suggest that in most physically interesting cases $I_1$ can be neglected. Thus, we propose the following simplification for the exact expression for the entropy (\ref{eqn:ent_sfin_2}) considering only leading order term
\begin{gather}
H \approx H_0 +  \sum_{k,m} \frac{1}{2\lambda_{k}(\lambda_k - \lambda_m)}{\sum_{l,n}(\psi^k_l)^2u_{ln}(\psi^m_n)^2}
\end{gather}

Nevertheless direct expression \ref{eqn:ent_sfin_2} still can be used in all desired cases instead of simplified version.





\end{document}